\documentclass[usenatbib]{mn2e}
\usepackage{graphicx}
\usepackage{natbib}
\usepackage{amsmath}
\def\apj{{ApJ}}                 % Astrophysical Journal
\def\apjl{{ApJ}}                % Astrophysical Journal, Letters
\def\aap{{A\&A}}                % Astronomy and Astrophysics
\def\mnras{{MNRAS}}             % Monthly Notices of the RAS
\def\prd{{Phys.~Rev.~D}}        % Physical Review D
\def\nat{{Nature}}              % Nature
\title[Wavelet analysis of 2dFGRS]
{The DWT Power Spectrum of the two-degree Field Galaxy Redshift Survey}

\begin{document}

\author[Cai et al.]{Yan-Chuan Cai$^{1,2}$, Jun Pan$^{1}$,
Yong-Heng Zhao$^{2}$,
 Long-Long Feng$^{1}$ and Li-Zhi Fang$^{3}$ \\
$^{1}$Purple Mountain Observatory, Chinese Academy of Sciences,
Nanjing 210008, China \\ $^{2}$National Astronomical Observatories,
Chinese Academy of Sciences,
Beijing 100012, China \\
$^{3}$Department of Physics, University of Arizona, Tucson, AZ 85721, US}

\maketitle
  
\begin{abstract}

The power spectrum of the two-degree Field Galaxy Redshift Survey (2dFGRS)
sample is estimated with the discrete wavelet
transform (DWT) method. The DWT power spectra within
$0.04 <k< 2.3 h$Mpc$^{-1}$ are measured for three
volume-limited samples defined in connective absolute
magnitude bins $-19 \sim -18$, $-20 \sim -19$ and $-21 \sim -20$.
We show that the DWT power spectrum can effectively distinguish
$\Lambda$CDM models of $\sigma_8=0.84$ and $\sigma_8=0.74$.
We adopt maximum likelihood method to perform three-parameter fitting with
bias parameter $b$, pairwise velocity dispersion $\sigma_{pv}$ and redshift distortion
parameter $\beta=\Omega_m^{0.6}/b$ to the measured DWT power spectrum.
Fitting results denotes that in a $\sigma_8=0.84$ universe the best fitted
$\Omega_m$ given by the three samples are consistent in the
range $0.28 \sim 0.36$, and the best fitted $\sigma_{pv}$ are
$398^{+35}_{-27}$, $475^{+37}_{-29}$ and $550 \pm 20$km/s for the three samples, respectively.
However in the model of $\sigma_8=0.74$, our three samples give very
different values of $\Omega_m$.
We repeat the fitting by using empirical formula of redshift distortion.
The result of the model of low $\sigma_8$ is still poor, especially, one of
the best value $\sigma_{pv}$ is as large as $10^3$km/s.
The power spectrum of 2dFGRS seems in disfavor of models with
low amplitude of density fluctuations.

\end{abstract}


%===================================================================
\section{Introduction}
The present clumpy structures indicated by galaxies on large scales
are evolved from the very small density fluctuations in the early era
of the universe. The amplitude of the fluctuation is fundamental to
understand the structure formation. A remarkable success of modern
cosmology is that the amplitude of mass fluctuations detected by the
anisotropy of cosmic microwave background radiation is in excellent
agreement with the analysis of the galaxy clustering at low
redshifts. Recently released WMAP third year data (WMAP3) refines most
results of cosmological parameters given by the WMAP 1st year data.
However, the fluctuation amplitude smoothed in a spherical top hat
window of radius of $8\rm{h^{-1} Mpc}$ is found as small as
$\sigma_8=0.74 ^{+0.05}_{-0.06}$ \citep{Spergel2006}, which is
significantly lower than $\sigma_8=0.84\pm 0.04$ of the WMAP 1st
year data. The new result of $\sigma_8$ is a challenge to the
cosmological parameter determinations from samples of galaxies and
galaxy clusters, most of which yield $\sigma_8 \simeq 0.9-1$ if the
matter content of the universe $\Omega_m \leq 0.3$
\citep[e.g.][]{2002ApJ...567..716R, 2002ApJ...577..595H,
2002ApJ...572L.131R,2002A&A...393..369V,2003MNRAS.344..673B,
2003ApJ...588L...1B, 2005PhRvD..71j3515S,2006MNRAS.365..231V}.

The problem motivates us to revisit the constrains
on $\sigma_8$ from the power spectrum of the sample of the two-degree
Field Galaxy Redshift Survey (2dFGRS). The final released
spectroscopic catalog of the 2dFGRS contains 221414
galaxies with good redshift quality $Q\geq 3$ and covers approximately 1800
square degrees of the sky. It is a good sample for studying the
fluctuations of cosmic mass field on large scales in linear regime
as well as on scales in nonlinear range. Moreover, the 2dFGRS
team has made detailed analysis of the Fourier power spectrum, the two
point correlation functions and relevant cosmological parameter fitting
\citep{2001MNRAS,2001MNRAS.328...64N, 2001Natur.410..169P,
2001MNRAS.328...64N, 2003MNRAS.346...78H, 2004MNRAS.353.1201P,
2005MNRAS.362..505C}. They found that $\Omega_m \sim 0.3$ or less
\citep{2001Natur.410..169P,2005MNRAS.362..505C}, and the best
fitting value of $\sigma_8$ is $\sim 0.95$ if one takes $\Omega_m
\sim 0.3$ \citep{2004MNRAS.353.1201P}, which is substantially
different from WMAP3. 

In the linear regime $\sigma_8$ only controls the
overall amplitude of the power spectrum, and it is degenerated with the 
linear bias parameter $b$. Power spectrum on small scales is more effective 
to constrain $\sigma_8$ than on large scales since the nonlinearity of the power spectrum
is directly reflected by the value of $\sigma_8$.
Most measurements of the 2dFGRS power spectrum are on scales
of $k < 1h$Mpc$^{-1}$ while cosmological parameter estimation is performed
on scale of $k<\sim 0.2h$Mpc$^{-1}$ \citep{2001Natur.410..169P,2001MNRAS,
2002MNRAS.335..887T,2005MNRAS.362..505C}. With the estimator based on the discrete 
wavelet transformation (DWT) 
\citep{2001PABei..19S..37Y, YangEtal2001a, 2001ApJ...560..549Y, 2002ApJ...566..630Y},
we are to analyze the power spectrum of the 2dFGRS sample on scales down to $k \simeq 2h$Mpc$^{-1}$. 
In this scale range, the redshift distortion of DWT diagonal mode power spectrum
can be easily approximated and the aliasing effect is exactly eliminated by the DWT
algorithm \citep{2000ApJ...539....5F}. 

The paper is organized as follows. In Section 2, the DWT power spectrum
estimator is briefly introduced. Section 3 describes the construction of samples.
The Section 4 demonstrates robustness and accuracy tests on 
the DWT power spectrum estimator. Section 5 lists our fitting results. 
Conclusions are stated in Section 6.

%==========================================================================
\section{DWT Power Spectrum Estimator}

The method of measuring galaxy power spectrum with the
multi-resolution analysis of discrete wavelet transformation has
been developed in the last decade \citep[e.g.][]{1995AAS...187.8502P,
1996ApJ...459....1P,2000ApJ...539....5F,
2001ApJ...560..549Y,2002ApJ...566..630Y,2003ApJ...585...12Z}.
A brief summary of the method is given here, more details are
written in the Appendix.

\subsection{DWT Power Spectrum}

The observed galaxy number density distribution is
%eq1
\begin{equation}
n_g({\bf x})=\sum_{m=1}^{N_g}w_m\delta^D({\bf x-x}_m)\ ,
\end{equation}
where $N_g$ is the total number of galaxies, ${\bf x}=(x_1,x_2,x_3)$
is a 3-dimensional position vector, ${\bf x}_m$ is the position of
the $m^{th}$ galaxy, $w_m$ is its weight, and $\delta^D$ is the 3-D Dirac
$\delta$ function. For an observed sample, $n_g({\bf x})$ can be
regarded as a realization of a Poisson point process with
intensity $n({\bf x})=\bar{n}_g({\bf x})[1+\delta({\bf x})]$, where
$\bar{n}_g({\bf x})$ is selection function, and $\delta({\bf x})$ is
the density contrast.

In terms of the DWT decomposition, the galaxy field is described
equivalently by variables defined as
%eq2
\begin{equation}
\tilde{\epsilon}_{\bf j, l}= \int \delta_g({\bf x})
      \psi_{\bf j,l}({\bf x})d{\bf x}\ ,
\end{equation}
where $\delta_g({\bf x})=[n_g({\bf x})/\bar{n}_g({\bf x})]-1$, and
$\psi_{\bf j,l}({\bf x})$ is the basis of the DWT decomposition,
where index ${\bf j}=(j_1,j_2,j_3)$ stands for the scale, and ${\bf
l}=(l_1,l_2,l_3)$ for position (see Appendix A). Since the
orthogonal-normal bases $\psi_{\bf j,l}({\bf x})$ are complete, all
second order statistical behavior of the field can be described by
$\langle \tilde\epsilon_{\bf j, l}\tilde\epsilon_{\bf j',
l'}\rangle$. The goal of power spectrum measurement is to estimate
the power spectrum of the density fluctuations $\delta({\bf
x})=[n({\bf x})/\bar{n}({\bf x})]-1$ from the observed realization
$\delta_g({\bf x})=[n_g({\bf x})/\bar{n}_g({\bf x})]-1$. It has been
shown in Fang \& Fang (2006, hereafter paper I) that the power of
the fluctuations on the modes with the scale index ${\bf j}$ can be
estimated by
%eq3
\begin{equation}
P_{\bf j}=I_{\bf j}^2-N_{\bf j}\ , 
\label{eq6}
\end{equation}
in which
%eq4
\begin{equation}
I_{\bf j}^2= \frac{1}{2^{j_1+j_2+j_3}}\sum_{l_1=0}^{2^{j_1}-1}
\sum_{l_2=0}^{2^{j_2}-1}\sum_{l_3=0}^{2^{j_3}-1}
  [\tilde\epsilon_{\bf j, l}]^2\ ,
\end{equation}
and
%eq5
\begin{equation}
N_{\bf j} = \frac{1}{2^{j_1+j_2+j_3}}\sum_{l_1=0}^{2^{j_1}-1}
\sum_{l_2=0}^{2^{j_2}-1}\sum_{l_3=0}^{2^{j_3}-1} \int\frac{\psi_{\bf
j,l}^2({\bf x})}{\bar{n}_g({\bf x})}
 d{\bf x}\ .
\end{equation}
The term $I_{\bf j}^2$ is the mean power of ${\bf j}$ modes measured
from the observed realization $n_g({\bf x})$, and $N_{\bf
j}$ is the power on ${\bf j}$ modes due to the Poisson noise. For a
volume-limited survey, the mean galaxy density $\bar{n}_g$ is
independent of the redshift. The Poisson noise power is thus simply
$1/\bar{n}_g$. $P_{\bf j}$ is usually referred to as the DWT power
spectrum.

The DWT power spectrum $P_{\bf j}$ is related to Fourier spectrum
$P(n_1,n_2,n_3)$ by
%eq6
\begin{equation}
\begin{aligned}
P_{\bf j} = & \frac{1}{2^{j_1+j_2+j_3}}
  \sum_{n_1 = - \infty}^{\infty}
  \sum_{n_2 = - \infty}^{\infty}
  \sum_{n_3 = - \infty}^{\infty}\\
  & |\hat{\psi}(n_1/2^{j_1})\hat{\psi}(n_2/2^{j_2})
  \hat{\psi}(n_3/2^{j_3})|^2 P(n_1,n_2,n_3)\ ,
\end{aligned}
\end{equation}
where $\hat{\psi}(n)$ is the Fourier transform of the basic wavelet
$\psi(x)$. Since $|\hat{\psi}(n)|^2$ is a high pass filter in the
wavenumber space, $P_{\bf j}$ is banded Fourier power spectrum. If
the cosmic density field is isotropic, the Fourier power spectrum
$P(n_1,n_2,n_3)$ depends only on $n=\sqrt{n_1^2+n_2^2+n_3^2}$.
Eq.(6) are exact for homogeneously random fields, either Gaussian or
non-Gaussian.

\subsection{DWT algorithm of redshift-distortion}

The DWT power spectrum depends on the scale and shape
of the DWT mode $\psi_{\bf j,l}({\bf x})$, it is sensitive to
distortion of the shape of the field. It is necessary to establish
the mapping from the redshift space to the real
space. The mapping is attributed to bulk velocity and pairwise
peculiar velocity. In the linear treatment of
bulk velocity, the redshift-distorted DWT power spectrum, $P^S_{\bf
j}$ is related to the real space power spectrum $P_{\bf j}$ by
\citep{2002ApJ...566..630Y}
%eq7
\begin{equation}
 P^S_{\bf j}=b^2(1+\beta S_{\bf j})^2 S^{PV}_{\bf j} P_{\bf j}\ ,
\end{equation}
in which $\beta=\Omega_m^{0.6}/b$, $b$ is the linear bias parameter. $S_{\bf j}$ 
of Eq.(7) is the linear redshift distortion factor. For a cubic
box of $L_1=L_2=L_3=L$, 
%eq8
\begin{equation}
\begin{aligned}
  S_{j_1,j_2,j_3}= &\frac{1}{2^{j_1+j_2+j_3}}
    \sum_{n_1,n_2,n_3 = \infty}^{\infty}
   \frac{n_3^2}{n_1^2 + n_2^2+ n_3^2} \\
   &\cdot|\hat{\psi}(n_1/2^{j_1})\hat{\psi}(n_2/2^{j_2})
    \hat{\psi}(n_3/2^{j_3})|^2\ .
\end{aligned}
\end{equation}
For diagonal modes $j_1=j_2=j_3=j$, $S_{j,j,j}=\frac{1}{3}$. 
The factor $S^{PV}_{\bf j}$ in Eq.(7) is
the pairwise velocity dispersion factor. In the plane-parallel approximation, if the
direction $j_3$ is chosen to be the line of sight, we have $S^{PV}_{\bf
j}=[s^{pv}_{\bf j}]^2$, with
%eq9
\begin{equation}
s^{pv}_{j_1,j_2,j_3} = \frac{1}{2^{j_3}}
   \sum_{n_3 = \infty}^{\infty}
 |\hat{\psi}(n_3/2^{j_3})|^2 \exp[-\frac{\sigma^2_{pv}}{2} (\frac{2\pi n_3}{L})^2]\ .
\end{equation}

Obviously, the factor $(1+\beta S_{\bf j})^2$ corresponds to the
linear redshift distortion, and $S^{PV}_{\bf j}$ is the nonlinear
redshift distortion caused by the pairwise velocity dispersion.
Although Eq.(7) is given by the linear approximation of bulk
velocity, N-body simulation points out that the mapping of Eq.(7) works well
till scale $k \simeq 2 h$Mpc$^{-1}$, also because $P_{\bf j}$ is weakly
affected by non-linear clumps of the density field. In general for non-volume
limited samples, selection function shall be
taken into account to model redshift distortion which brings in high order correction to
Eq.(7) \citep{2002ApJ...566..630Y}.

%========================================================================================
\section{Sample Construction}
\begin{table*}
\centering
\caption{Volume limited sub-samples of 2dFGRS.}
\bigskip
\begin{tabular}{lccccccc}
\hline
$M_{b_J}-5\log_{10} h$ & $z_{min}$ & $z_{max}$ & $d_{min}$ & $d_{max}$ &
   $N_g^{SGP}/N_g^{NGP}$ &
 $\bar n(10^{-3}h^3{\rm Mpc}^{-3})$ \\
\hline
-19 --- -18 & 0.0205 & 0.087  & 61.2  & 255.7  &  9737/7811   &  8.393 \\
-20 --- -19 & 0.0320 & 0.129  & 95.2  & 374.9  &  19122/14390 &  5.102 \\
-21 --- -20 & 0.0495 & 0.186  & 146.6 & 532.9  &  14734/10202 &  1.330 \\
\hline
\end{tabular}
\label{tb:2dfdata}
\end{table*}
Samples used in our analysis are constructed basically in the
same way as that in \cite{2005MNRAS.362.1363P}. We create
volume limited samples from the 2dFGRS spectroscopic catalog of
the final data release \citep{2003yCat.7226....0C}, which contains
221414 galaxies with good redshift quality $Q\ge3$
\citep{2001MNRAS.328.1039C}. We exclude the ancillary random fields,
leaving the two major contiguous trunks: one near the South Galactic
Pole (SGP) covering approximately
$-37^\circ\negthinspace.5<\delta<-22^\circ\negthinspace.5$, $21^{\rm
h}40^{\rm m}<\alpha<3^{\rm h}40^{\rm m}$ and the other around the
North Galactic Pole (NGP) defined roughly by
$-7^\circ\negthinspace.5<\delta<2^\circ\negthinspace.5$, $9^{\rm
h}50^{\rm m}<\alpha<14^{\rm h}50^{\rm m}$. 
In order to get maximum
number of galaxies while keep an uniform sampling rate to guarantee
the fairness of our statistics, we try different values of
completeness $f$ ($f$ is defined as the ratio of the number of
galaxies with redshift to the total number
of galaxies contained in the parent catalog): fields with
completeness less than the chosen value is cut off, and fields with
higher completeness is diluted to match the sampling rate. We find
$f=0.738$ is the optimal value. The final parent samples are
thus restricted by completeness $f>0.738$, and apparent
magnitudes limits in photometric $b_J$ band with bright cut of
$m_{b_J}=15$ and faint cut of median value of $\sim19.3$ with
some small variation specified by masks \citep{2003yCat.7226....0C}.

Volume limited sub-samples are built from the parent sample by
selecting galaxies in specified absolute magnitude ranges. Absolute
magnitudes are calculated with $k+e$ correction in
\cite{2002MNRAS.336..907N}. Our analysis focus
on the three sub-samples defined in absolute magnitude $M_{b_J}$ bins of 
$-19 \sim -18$, $-20 \sim -19$ and $-21 \sim -20$. Basic parameters of the three volume
limited samples of 2dFGRS are summarized in Table~\ref{tb:2dfdata}
in which lists range of redshifts $z_{min}$-$z_{max}$, range of comoving distances $d_{min}$-
$d_{max}$,  numbers of SGP and NGP galaxies, and the mean densities $\bar{n}$. 
Comoving distances are calculated from redshifts $z$ in the $\Lambda$CDM universe with
$\Omega_\Lambda=0.7$ and $\Omega_m=0.3$. 

%==========================================================================
\section{Numerical Tests of the DWT Power Spectrum Estimator}
\begin{figure}
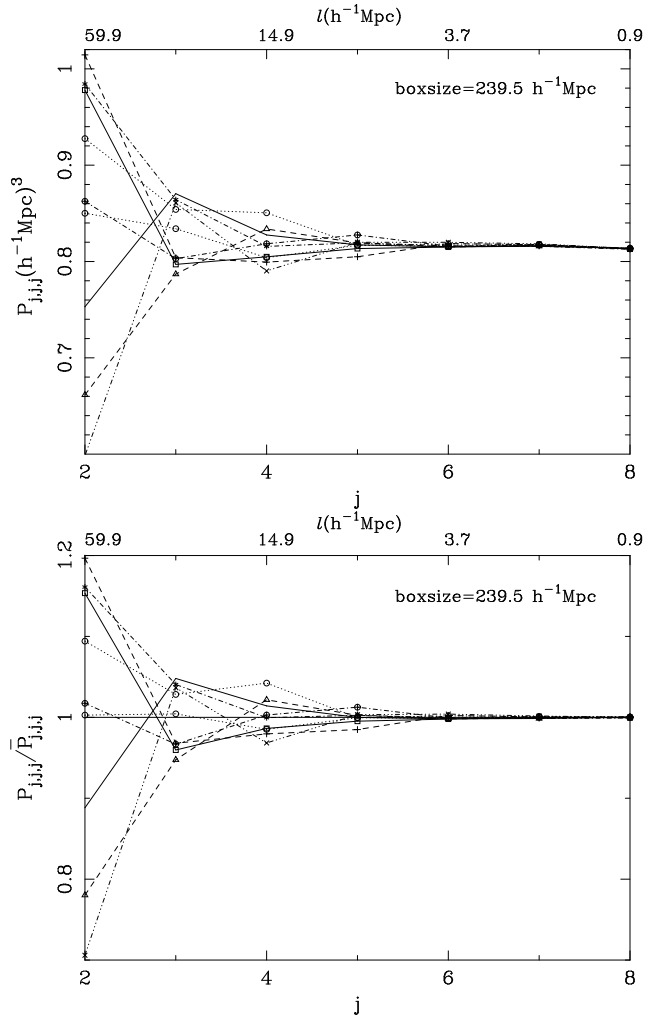

\resizebox{\hsize}{!}{
\includegraphics[angle=270]{random_pow.ps}}
\resizebox{\hsize}{!}{
\includegraphics[angle=270]{random_ratio.ps}}
\caption{$Top$: Diagonal DWT power spectra of nine realizations of
Poisson samples. Each sample is generated in a box with size of
$239.5$ h$^{-1}$Mpc and $256^3$ particles. The relation between
$j$ and physical scale is $l=239.5/2^j$ h$^{-1}$Mpc. $Bottom$: Ratios of
diagonal DWT power spectra of nine realizations of Poisson samples
to their mean spectrum.
\label{fig-1}}
\end{figure}

In this section, we will test the DWT power spectrum estimator with
Poisson samples and samples from N-body simulation.
Nine realizations of Poisson samples are produced in box with side
$L=239.5$ h$^{-1}$Mpc with $256^3$ points.
All the measured DWT power spectra of these samples are plotted in the
top panel of Figure \ref{fig-1}. To test the stability of the DWT
estimator we calculate the diagonal DWT power spectrum, $P_{j,j,j}$
for each realization, and then computing their mean
$\overline{P}_{j,j,j}$. $P_{j,j,j}/\overline{P}_{j,j,j}$
are shown in the bottom panel of Figure \ref{fig-1}. We can see that
at $j\le 2$(i.e. scales larger than $119.75 h$$^{-1}\rm{Mpc}$),
there are as large as 50\% variances in the diagonal DWT power
spectrum. Thus we will not use data points with  $j\le 2$. At
small scales, or large $j$, the DWT estimator gives reliable
results. This is because the aliasing effect is effectively
suppressed in the DWT analysis \citep{2000ApJ...539....5F}.

\begin{figure}
\resizebox{\hsize}{!}{
\includegraphics[angle=270]{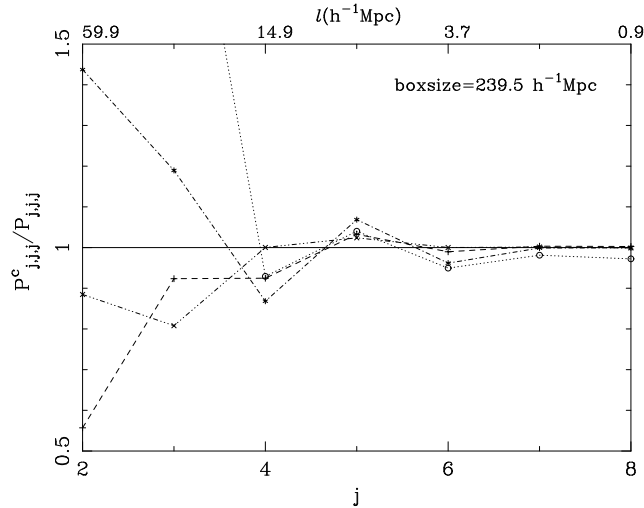}}
\caption{The ratios of DWT power spectra of cutted Poisson sub-samples to
the parent Poisson sample. The parent Poisson sample is generated in a
box with size of $239.5$ h$^{-1}$Mpc, with $256^3$ points.
The sub-samples are defined in box of $60.0\times239.5\times239.5 h^{-1}$Mpc
(dash-dot-dot line); $20.0\times239.5\times239.5 h^{-1}$Mpc (dash line);
$20.0\times60.0\times239.5 h^{-1}$Mpc (dash-dot line);
and $20.0\times60.0\times239.5 h^{-1}$Mpc with three cylinders chopped
(dot line).
\label{fig-2}}
\end{figure}

Then, in order to test the geometric effect of samples on
estimation of the power spectrum, we cut one of the Poisson sample
into three sheet-like sub-samples as $60.0\times239.5\times239.5$,
$20.0\times239.5\times239.5$, and $20.0\times60.0\times239.5$
(h$^{-1} \rm{Mpc})^3$. Furthermore, a fourth sub-sample is constructed from the
$20.0\times239.5\times239.5$ (h$^{-1}\rm{Mpc})^3$ sub-sample by cutting
off three parallel cylinders with radius of 5.0, 10.0 and 20.0 $h^{-1}\rm{Mpc}$ 
respectively. In Figure \ref{fig-2}, we plot the ratios of the DWT power spectra of
each sub-sample to their parent random sample
$P^c_{j,j,j}/P_{j,j,j}$. For $j=2$ and 3 (or on the scale of
$239.5/2$ h$^{-1}\rm{Mpc}$ and $239.5/2^2$ h$^{-1} \rm{Mpc}$), the
scatters in those spectra can be larger than $50\%$, and the ratios
are randomly distributed with variance of the order of unity.
Actually, such significant scatters result from small number of
modes on $j \leq 3$. For $j>3$, differences between those
spectra are negligible, the DWT power spectrum estimator is well
independent of sample geometry on small scales.

\begin{figure}
\resizebox{\hsize}{!}{
\includegraphics[angle=270]{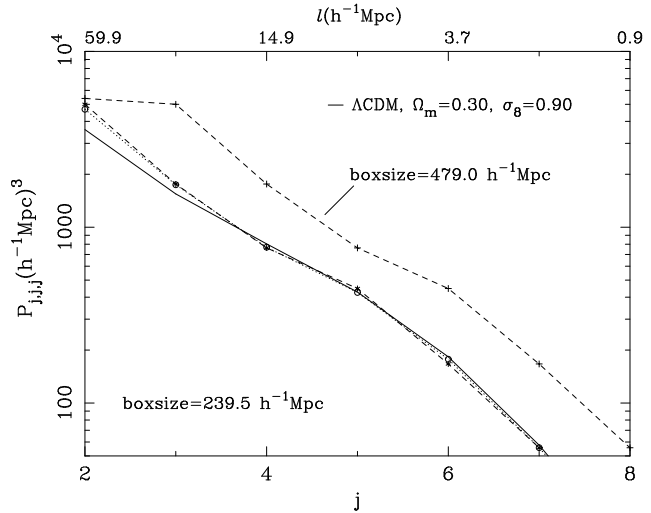}}
\caption{DWT power spectrum of Virgo simulation sample. The dash
line on the right is measured with the box size of 479.0
$h^{-1} \rm{Mpc}$. Shifting the right dash line by one unit along
j-axis, we get the dash line on the left. The dot line is
the power spectrum measured in box of size 239.5
$h^{-1} \rm{Mpc}$. The solid line is the nonlinear power spectrum
from the formula of \citet{2003MNRAS.341.1311S} with $\Omega_m=0.3$, 
$\Omega_{\Lambda}=0.7$ and $\sigma_8=0.9$.
\label{fig-3}}
\end{figure}

To further test the reliability of the DWT power spectrum estimator,
we measured the DWT power spectrum of the Virgo simulation which is of
$\Lambda$CDM cosmology with $256^3$ particles in box of size
$239.5 h^{-1}$Mpc. Figure \ref{fig-3} compares the theoretical DWT power 
spectrum with that estimated from the simulation. The theoretical DWT power 
spectrum is calculated from Eq.(6) with nonlinear power spectrum from the 
accurate fitting formula of \citet{2003MNRAS.341.1311S}. Clearly, the 
theoretical spectrum and measurements are in good agreement on scales 
less than $239.5/2^4 \simeq 15h^{-1}$Mpc. The test shows that the DWT 
power spectrum estimator can perfectly recover the original power spectrum 
on small scales.

The final test is to measure the DWT spectrum of the Virgo sample in a cubic 
window of side $479.0h^{-1}$Mpc which is twice of the simulation box size. 
Theoretically the power at scale $j$ in the box of side $479.0 h^{-1}$Mpc
corresponds to that at scale $j-1$ in the box of side $239.5h^{-1}$Mpc. 
It is clearly seen in Figure \ref{fig-3} that the spectrum measured
in the $479.0h^{-1}$Mpc box at $j$ exactly equals to that measured in the
$239.5 h^{-1}$Mpc at $j-1$. The DWT estimator is independent of the size of the 
window box. One can choose the size of box freely and put the sample in the box 
wherever one like. Note that the difference of spectra at $j=2,\ 3$ are
greater than other data points. Therefore, in the following analysis
of the 2dFGRS catalogs, the first two data points in our spectra are discarded.

%=========================================================================
\section{DWT power spectrum of 2dFGRS samples}

\subsection{The diagonal DWT power spectrum}
To achieve the largest possible volume, the NGP and the SGP regions 
are measured together in a window box of size $1403.0h^{-1}$Mpc. The size of the window
is sufficient to cover all three volume limited sub-samples. 
The filling factor of each sub-sample is $0.07\%$, $0.24\%$, and $0.68\%$
respectively. 

To estimate error bars of the DWT power spectra, we created mock
volume limited samples from the 22 mock galaxy catalogs which are
extracted from the Hubble volume simulation 
\footnote{http://star-www.dur.ac.uk/$\sim$cole/mocks/hubble.html}. 
Details of mock catalogs are in \citet{1998MNRAS.300..945C}. The set of
the mock samples used here is the LambdaCDM04. The 22 mock samples 
are filtered with the same selection criteria and masks as the real galaxy 
volume limited sub-samples. Error bars of the DWT
power spectra of the three 2dFGRS sub-samples are
approximated by the 1-$\sigma$ scattering of their mock samples.

Short noise is not directly calculated with Eq.(5). Instead, we
produce a number of Poisson samples with the same geometry as our
samples. Numbers of points in these Poisson samples are the same as 
the number of galaxies in the sample under analysis. 22 Poisson samples 
for each sub-sample allow us to estimate error bars for short noise subtraction. 
These error bars are incorporated into the final results of galaxy power
spectra.
\begin{figure}
\resizebox{\hsize}{!}{
\includegraphics[angle=270]{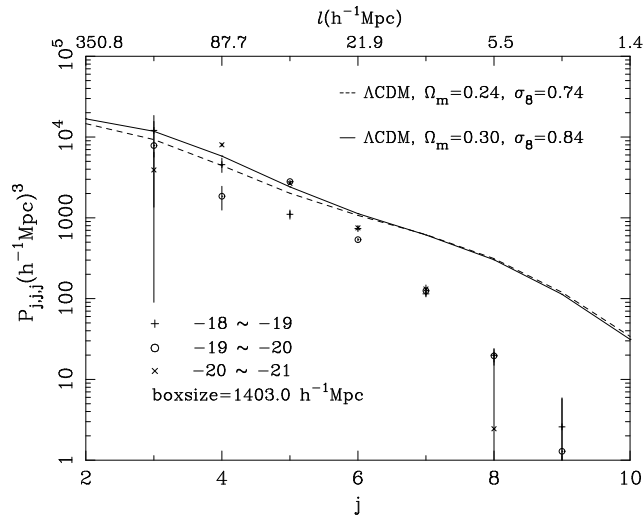}}
\caption{DWT power spectrum of 2dFGRS volume-limited samples and
theoretical spectra. The size of the cubic box is 1403.0
$h^{-1} \rm{Mpc}$. The relation between $j$ and physical scale is
$l=1403.0/2^j$ h$^{-1}$Mpc. The power at large scales are suppressed
by random
motion of galaxies. Due to cosmic variance, the error bars at $j=3$
is large. At $j=9$, Poisson noise leads to large error bars.
\label{fig-4}}
\end{figure}

In Figure \ref{fig-4} we present the measured diagonal DWT power
spectra of the three volume limited samples, together with
two theoretical nonlinear spectra of flat $\Lambda\rm{CDM}$ model with parameters
A.) $\Omega_m=0.3$, $\Omega_{\Lambda}=0.7$ and $\sigma_8=0.84$ (Model A), 
and B.) $\Omega_m=0.24$, $\Omega_{\Lambda}=0.76$ and $\sigma_8=0.74$ (Model
B). The scale range is $0.035 < k < 2.2h$Mpc$^{-1}$.

\subsection{The fitting of redshift distorted power spectrum}
The differences between the power spectra of model predictions and
the real data shown in Figure \ref{fig-4} are mainly due to 
redshift distortion and bias. We adopt Eq. (7) to fit the power spectrum 
of 2dFGRS samples with the nonlinear real space DWT power spectrum 
from Eq.~(6) and the formula of \citet{2003MNRAS.341.1311S}. We
firstly use the 22 mock samples to estimate the correlation between
the powers of different DWT modes. We construct a covariance matrix
%eq10
\begin{equation}
\tilde C_{jj'} = \frac{1}{N_{sim}}\sum_{I=1}^{N_{sim}} \Delta d^I_j
\Delta d^I_{j'}\ ,
\end{equation}
where $N_{sim} = 22$ and $\Delta d^I_j = d^I_j - \langle d_j
\rangle$. The vector $d^I$ consists of elements $d^I_j=P^I_{j,j,j}$
with $j=3,...,9$, $P^I_{j,j,j}$ is the power spectrum from the
$I$-th simulation, and $\langle d_j \rangle$ is the mean. We found
that off-diagonal elements of the covariance matrices are always one
order of magnitude smaller than diagonal elements. Actually it is a
typical feature in the DWT analysis. The correlations between
different modes are highly suppressed regardless the field is
Gaussian or non-Gaussian (Feng \& Fang 2005). The
quasi-diagonalizing of the correlation matrix in the DWT
decomposition has been extensively used for data compression. By
virtue of this property, we can compute $\chi^2$ with
diagonal elements only. Namely, we use the Chi-Square as our maximum
likelihood estimator
%eq11
\begin{equation}
\chi^2=\sum^{N}_{j=1}\frac{[P_j-P^S_j(a_1,a_2,a_3...)]^2}{\sigma_j^2}\ ,
\end{equation}
in which $P_j$ is observation data and $P^S_j(a_1,a_2,a_3...)$ is the
redshift-distorted power of model with parameters
$(a_1,a_2,a_3...)$. We take the Reduced-Chi-Square, which is defined
as $\chi^2_{d.o.f}=\chi^2/(N-M)$ where $(N-M)$ is the degree of
freedom, as our final results shown in tables.

We aim at detecting the influence of $\sigma_8$ on the power spectrum.
Two fiducial $\Lambda$CDM models are considered here: A. $\sigma_8=0.84$, and B.
$\sigma_8=0.74$. We select the linear bias parameters $b$, redshift
distortion parameter $\beta$ (or $\Omega_m$) and pairwise velocity
variance $\sigma_{pv}$ as fitting parameters. Other parameters of model A and B 
are the same. The parameter space $(b,\beta,\sigma_{pv})$ is
divided into a $20\times 20\times 20$ grid. The first run of fitting is 
performed on very crude grids in broad parameter space to locate the region of best $\chi^2$. 
Then we decrease the volume of parameter space centered in this region 
with finer grid to obtain the three dimensional probability
distribution functions (PDF) of $(b,\beta,\sigma_{pv})$. After
integrating over two of the three axes in the parameter space, we
have the marginalized PDF for each parameter.

\subsection{Fitting results}

\begin{table*}
\centering \caption{Parameters estimated by fitting 2dFGRS DWT power spectra with Eq.~(7)
to $\Lambda\rm{CDM}$ model of $\sigma_8=0.84$.}
\begin{tabular}{lcccc}
\hline
$M_{b_J}-5\log_{10} h$ & $\beta$ /$\Omega_m$ & $b$ & $\sigma_{pv} $ ($\rm{km s^{-1}}$) & $\chi^2$\\
\hline -19 --- -18  & $ 0.66^{+ 0.06}_{- 0.10}$ /$0.31^{+0.08}_{-0.12}$ & $ 0.75^{+ 0.05}_{- 0.06}$
  & $ 398^{+ 36}_{- 27} $  &  8.87 \\
-20 --- -19  & $ 0.62^{+ 0.07}_{- 0.10}$ /$0.36^{+0.12}_{-0.14}$ & $ 0.86^{+ 0.08}_{- 0.07}$
  & $ 475^{+ 37}_{- 29} $ &  16.19 \\
-21 --- -20  & $0.43^{+ 0.02}_{- 0.02}$ /$0.28^{+0.02}_{-0.02}$ &
$ 1.07^{+ 0.01}_{- 0.01}$
  & $ 550^{+ 20}_{- 20} $ &  4.06 \\
\hline
\end{tabular}
\label{tb:3p_DWT_sig0.84}
\end{table*}

\begin{table*}
\centering \caption{Parameters estimated by fitting 2dFGRS DWT power spectra with Eq.~(7)
to $\Lambda\rm{CDM}$ model of $\sigma_8=0.74$.}
\begin{tabular}{lcccc}
\hline
$M_{b_J}-5\log_{10} h$ & $\beta$ /$\Omega_m$ & $b$ & $\sigma_{pv}$ ($\rm{km s^{-1}}$) & $\chi^2$\\
\hline -19 --- -18  & $ 0.76^{+ 0.07}_{- 0.09} $
/$0.41^{+0.10}_{-0.12}$ & $ 0.77^{+ 0.05}_{- 0.05}$ & $ 415^{+ 40}_{- 26} $  &  12.24 \\
-20 --- -19  & $ 0.73^{+ 0.12}_{- 0.10} $ /$0.46^{+0.20}_{-0.16}$ &
$ 0.86^{+ 0.09}_{- 0.07}$
  & $ 492^{+ 38}_{- 40} $ & 17.99  \\
-21 --- -20  & $0.35^{+ 0.02}_{- 0.02} $ /$0.28^{+0.02}_{-0.02}$ &
$ 1.19^{+ 0.01}_{- 0.01}$
  & $ 600^{+ 15}_{- 15} $ &  4.83 \\
\hline
\end{tabular}
\label{tb:3p_DWT_sig0.74}
\end{table*}

\begin{table*}
\centering \caption{Parameters estimated by fitting 2dFGRS DWT power spectra with Eq.~(7)
to the model of $\Omega_m=0.24$ and $\Omega_{\Lambda}=0.76$.}
\begin{tabular}{lcccc}
\hline
$M_{b_J}-5\log_{10} h$ & $\sigma_8$ & $b$ & $\sigma_{pv}$ $\rm{km s^{-1}}$ & $\chi^2$\\
\hline
-19 --- -18 & $ 0.43^{+0.20}_{-0.06}$ & $0.99^{+ 0.20}_{- 0.10}$ & $ 490^{+ 42}_{- 40}$ &  10.74 \\
-20 --- -19  & $ 0.97^{+0.09}_{-0.06}$  & $ 0.82^{+ 0.15}_{- 0.13}$ & $445^{+ 45}_{- 35}$ &  16.83 \\
-21 --- -20  & $ 0.94^{+0.04}_{-0.04}$  & $ 0.99^{+ 0.09}_{- 0.05}$ & $505^{+ 40}_{- 35}$ &  4.77 \\
\hline
\end{tabular}
\label{tb:DWT_sig8_fit_Om_0.27}
\end{table*}

\begin{figure}
\resizebox{\hsize}{!}{
\includegraphics{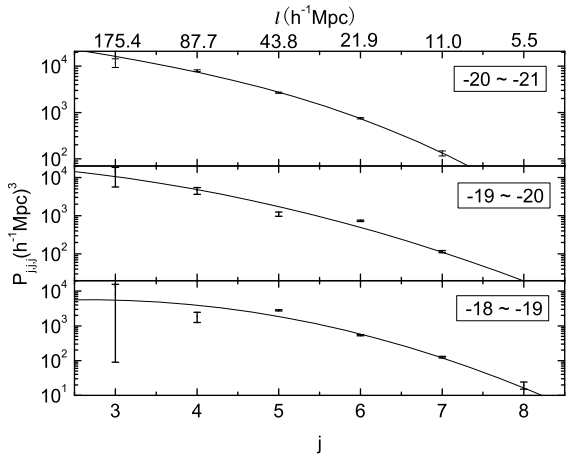}}
\caption{The DWT power spectra in redshift space of the
$\rm{\Lambda CDM}$ model with $\sigma_8=0.84$. Parameters
$b$, $\beta$, and $\sigma_{pv}$ are given by our best fitting
to the data points shown in the Figure. The three
panels are for sample within absolute magnitudes $-20\sim-21$ (top); 
$-19\sim -20$ (middle) and $-18 \sim -19$ (bottom).
\label{fig-5}}
\end{figure}

\begin{figure*}
\resizebox{\hsize}{!}{
\includegraphics{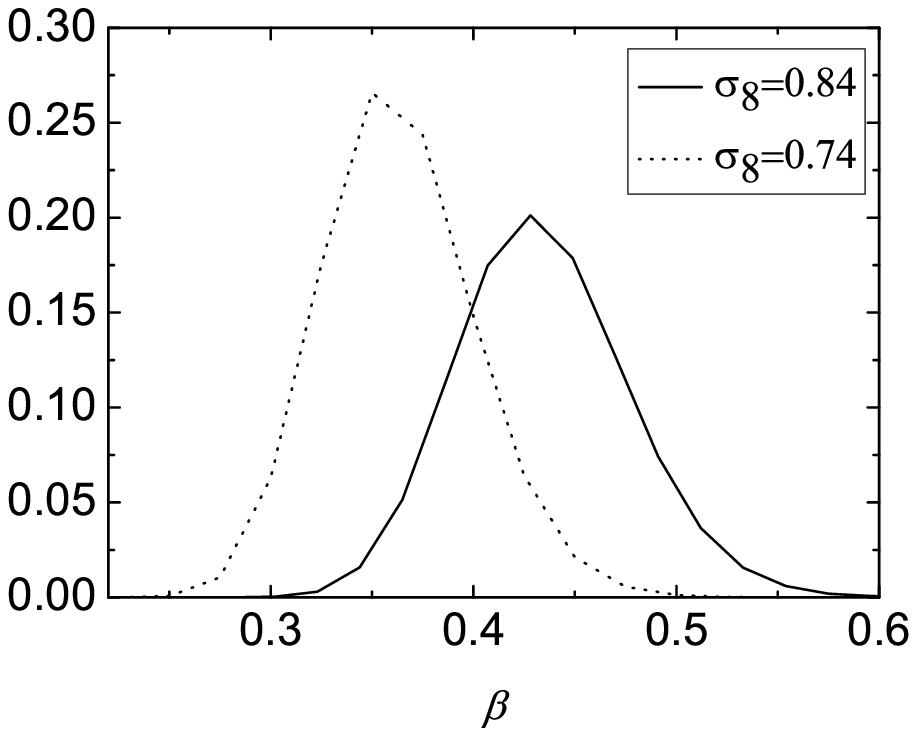}
\includegraphics{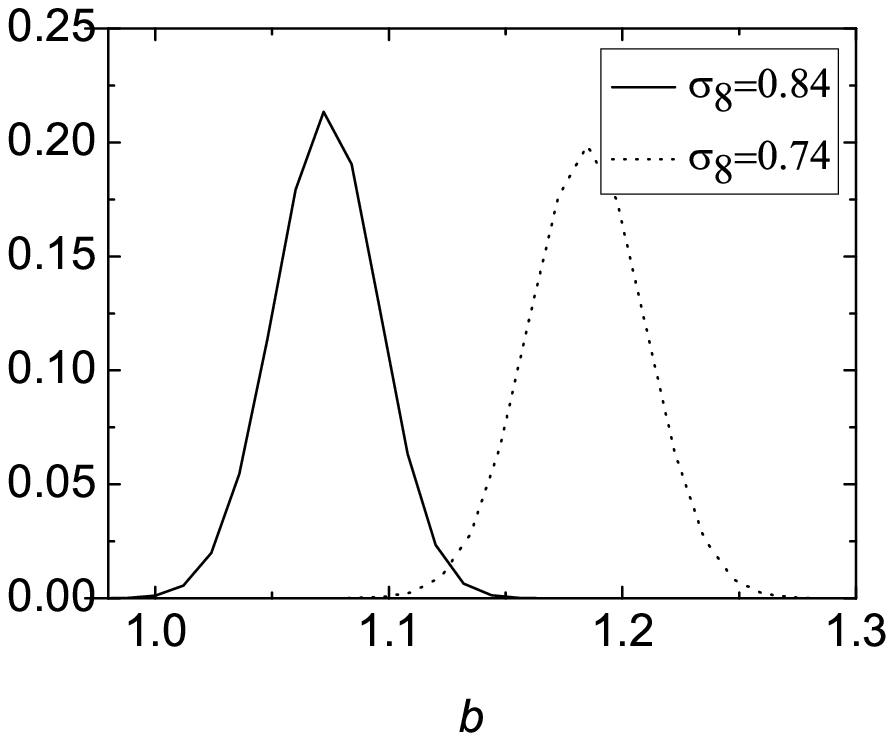}
\includegraphics{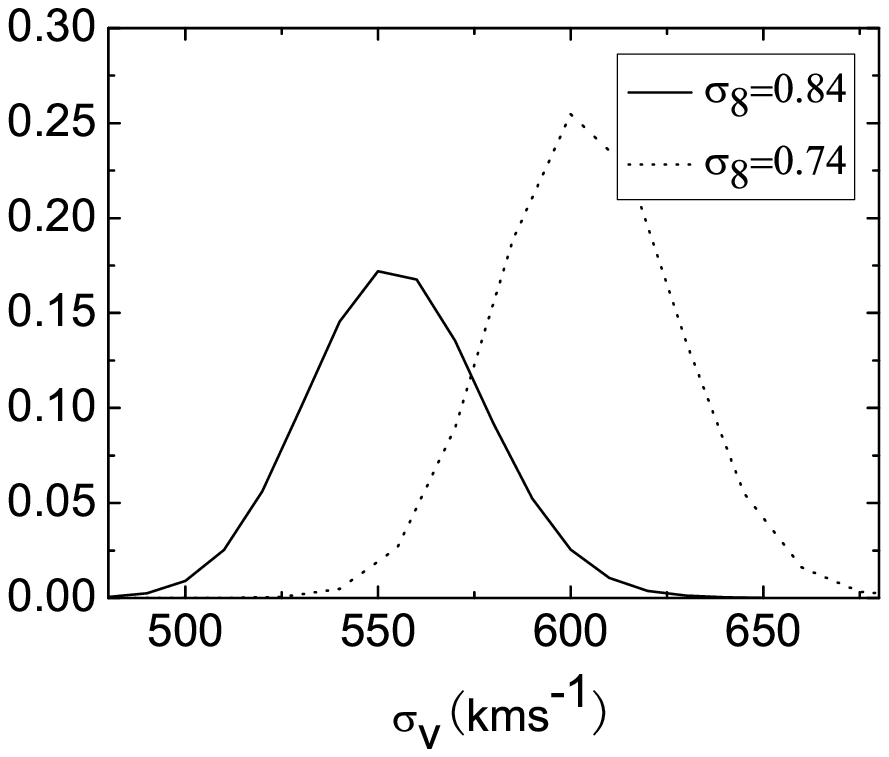}}
\caption{Marginalized distribution of parameters $b$, $\beta$, and $\sigma_{pv}$ in
fitting the DWT power spectrum of the 2dFGRS sample of 
$M_{b_J}\in(-20 \sim -21)$ with Eq.~(7).
\label{fig-6}}
\end{figure*}

\begin{figure*}
\resizebox{\hsize}{!}{
\includegraphics{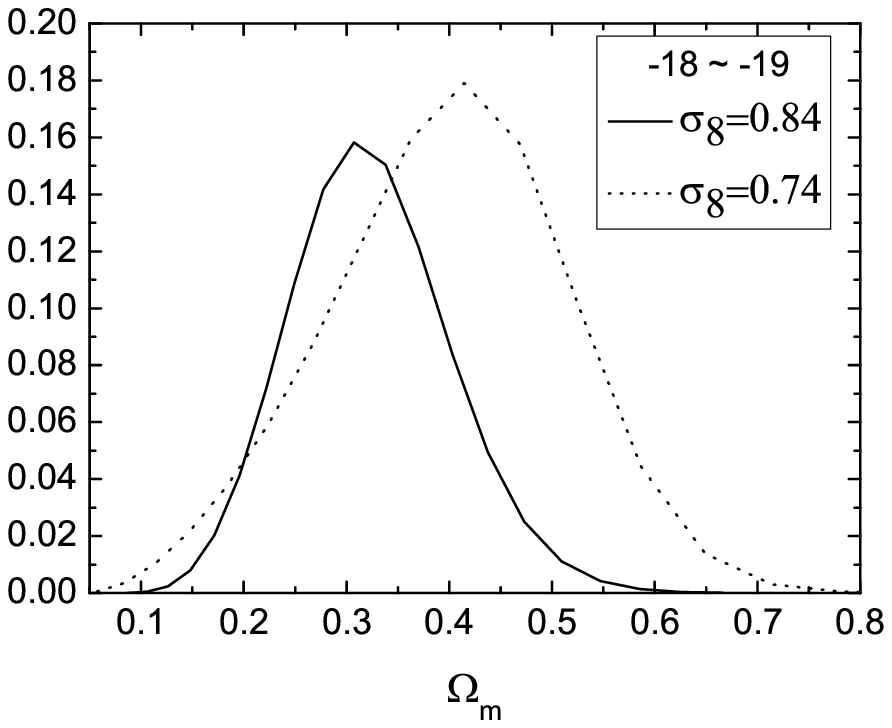}
\includegraphics{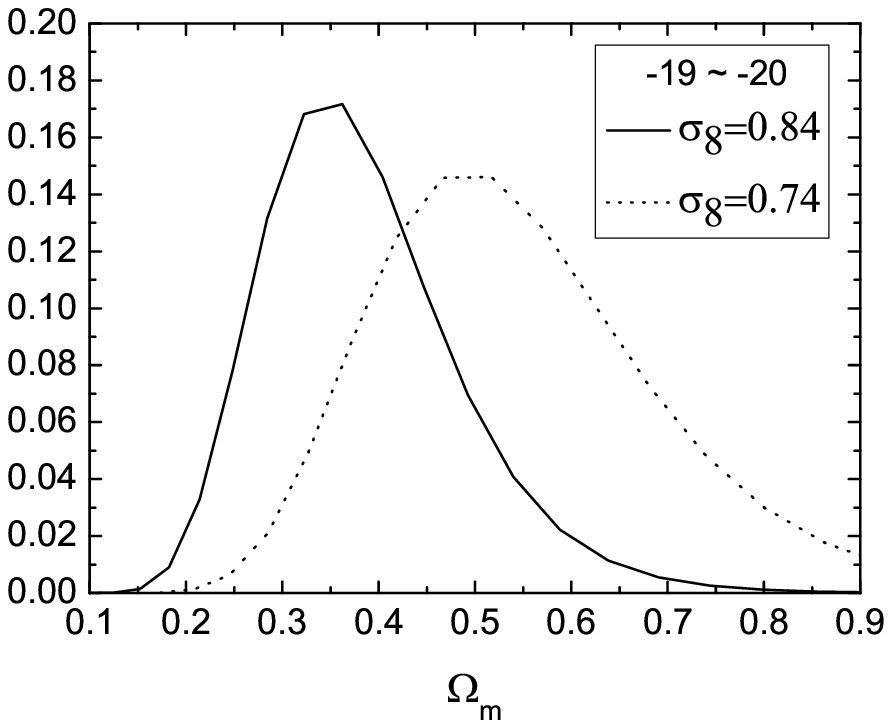}
\includegraphics{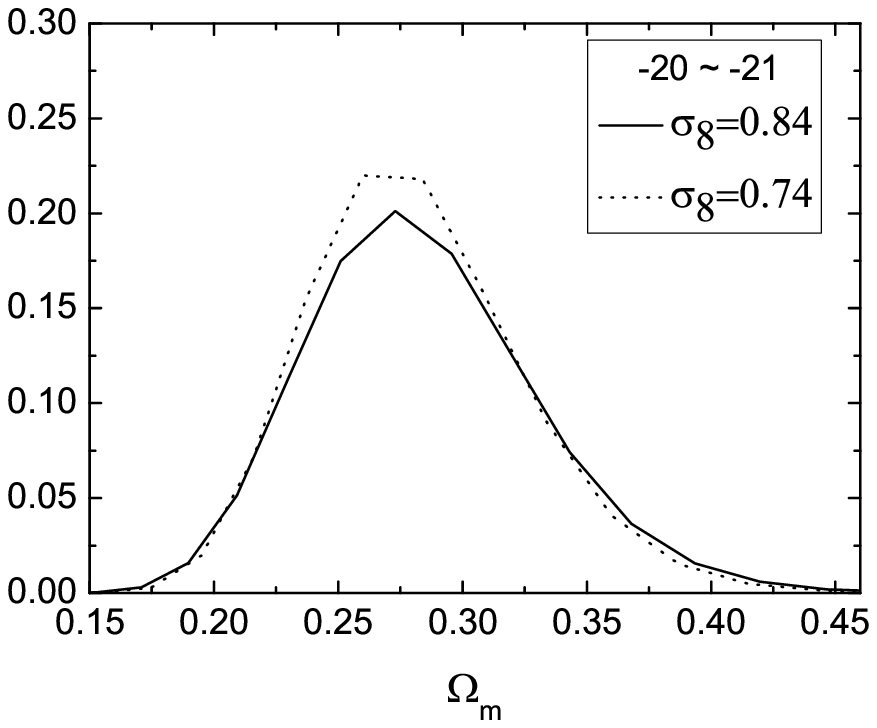}}
\caption{Marginalized distributions of $\Omega_m$ in fitting
the DWT power spectra with Eq.~(7) for 2dFGRS samples with $M_{b_J}$ in
$-18 \sim 19$, $-19 \sim -20$ and $-20 \sim -21$.
\label{fig-7}}
\end{figure*}

The best fitted power spectra of both models are very close to each
other, so only the fitted power spectra of model A are demonstrated 
in Figure \ref{fig-5}. The estimated parameters of the model A and B are
tabulated in Tables \ref{tb:3p_DWT_sig0.84} and
\ref{tb:3p_DWT_sig0.74} respectively. As an example, marginalized PDFs of the 
parameters $b, \beta, \sigma_{pv}$ for the sub-sample $-21 \sim -20$ are 
in Figure \ref{fig-6}. Figure \ref{fig-7} gives the PDFs of $\Omega_m$ for all the three
sub-samples.

In Table \ref{tb:3p_DWT_sig0.84}
we can see that three sub-samples offer about the same estimation of 
$\Omega_m\simeq 0.3$, and the $b$ increases from $0.75$ to $\simeq 1$ with 
the luminosity of galaxies, which is consistent with analysis of others
\citep{2001MNRAS.328...64N, 2005MNRAS.362.1363P}. The
pairwise velocities obtained in the three samples are in the
reasonable range, and also increase with galaxy luminosity.
It appears that our analysis with model A ($\sigma_8=0.84$) is basically
in good agreement with previous works, and more importantly, that the DWT
proves itself an effective tool for parameter estimation from galaxy samples.

Fitting to model B provides very different estimation of parameters. As seen in
Table \ref{tb:3p_DWT_sig0.74} and Figure \ref{fig-7}, values of $\Omega_m$ given by the
three sub-samples are quite different from each other. The best
values of $\Omega_m$ for the sub-samples of $-19 \sim -18$ and $-20 \sim
-19$ are significantly larger than $0.3$, which is in disagreement with
most current measurements at least at 1-$\sigma$ level
\citep[e.g.][]{2001Natur.410..169P,2004PhRvD..69j3501T}. Only the
sub-sample of $-21 \sim -20$ yields the $\Omega_m\approx 0.3$.
 
In order to place constrains from the 2dFGRS on $\sigma_8$, we repeat the
fitting procedure with $\sigma_8$, $b$ and
$\sigma_{pv}$ as fitting parameters, and $\Omega_m=0.24$ as prior. 
Results are shown in Table~\ref{tb:DWT_sig8_fit_Om_0.27}. It is very 
clear that, except the $-19 \sim -18$ sub-sample, the other two sub-samples both give
$\sigma_8>0.9$. An universe of $\sigma_8=0.74$ is not preferred by the
DWT power spectrum of the 2dFGRS.

%-----------------------------------------------------------------------------
\subsection{Fitting with an alternative formula of redshift distortion}
\begin{figure*}
\resizebox{\hsize}{!}{
\includegraphics{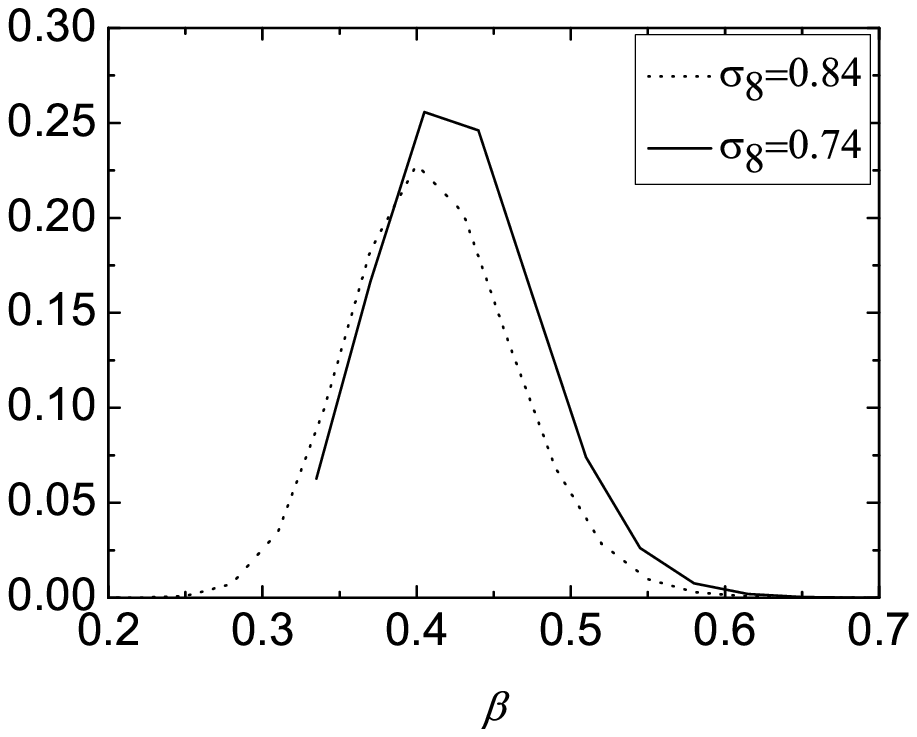}
\includegraphics{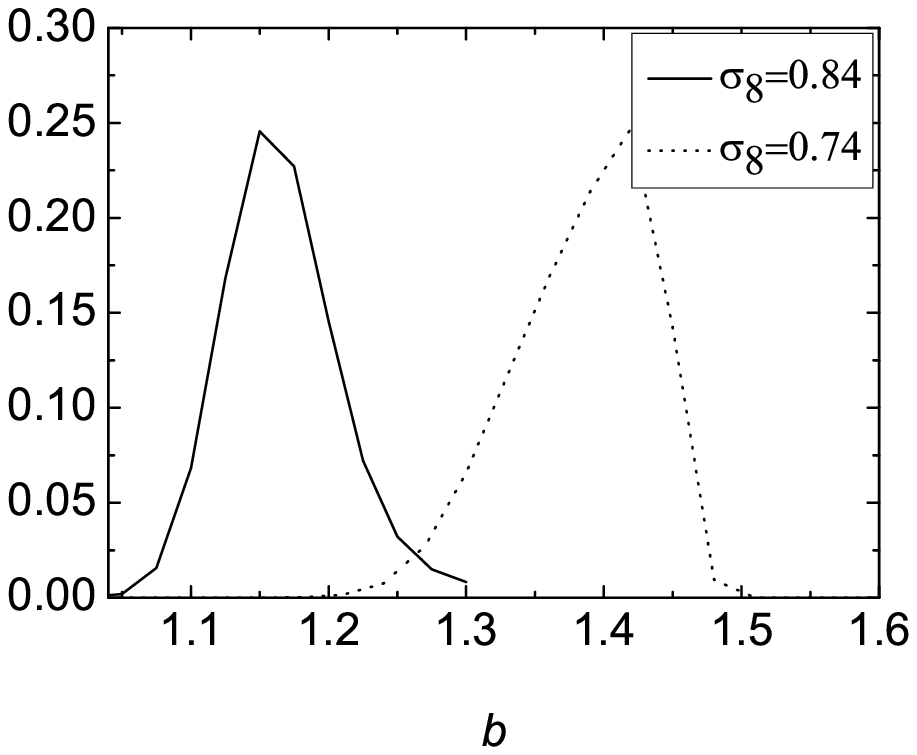}
\includegraphics{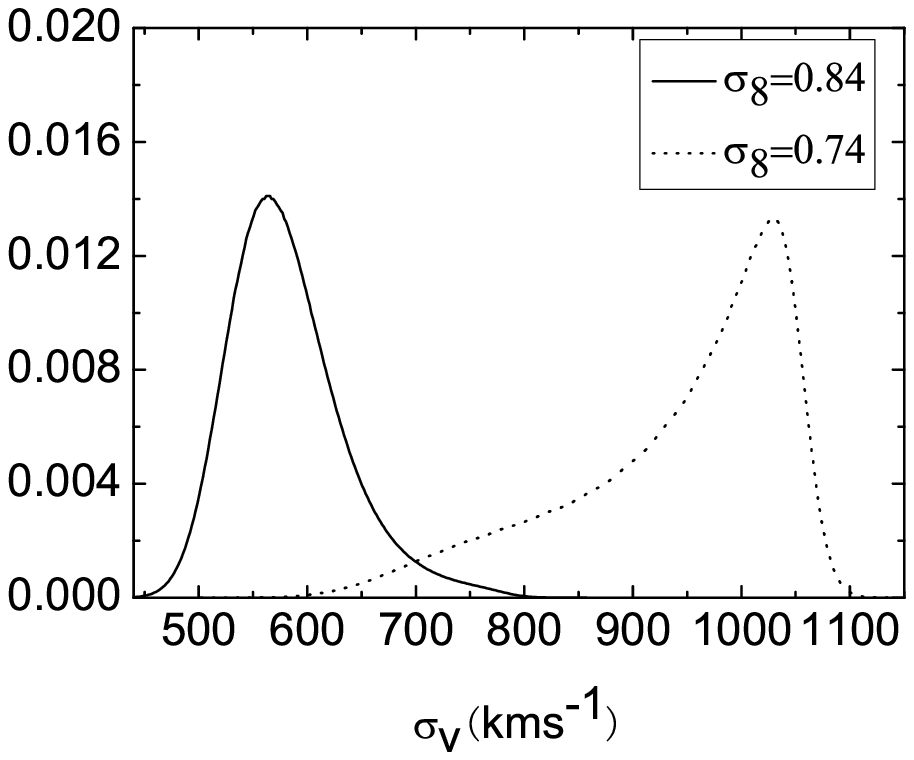}}
\caption{Marginalized distribution of parameters $\beta, b, \sigma_{pv}$ for the 
$-20 \sim -21$ sub-sample, with the empirical redshift distortion mapping formula of 
Eq.~(12). Fitting with $\sigma_8=0.74$ give larger $b$ and $\sigma_{pv}$
than with $\sigma_8=0.84$.
\label{fig-8}}
\end{figure*}

\begin{figure*}
\resizebox{\hsize}{!}{
\includegraphics{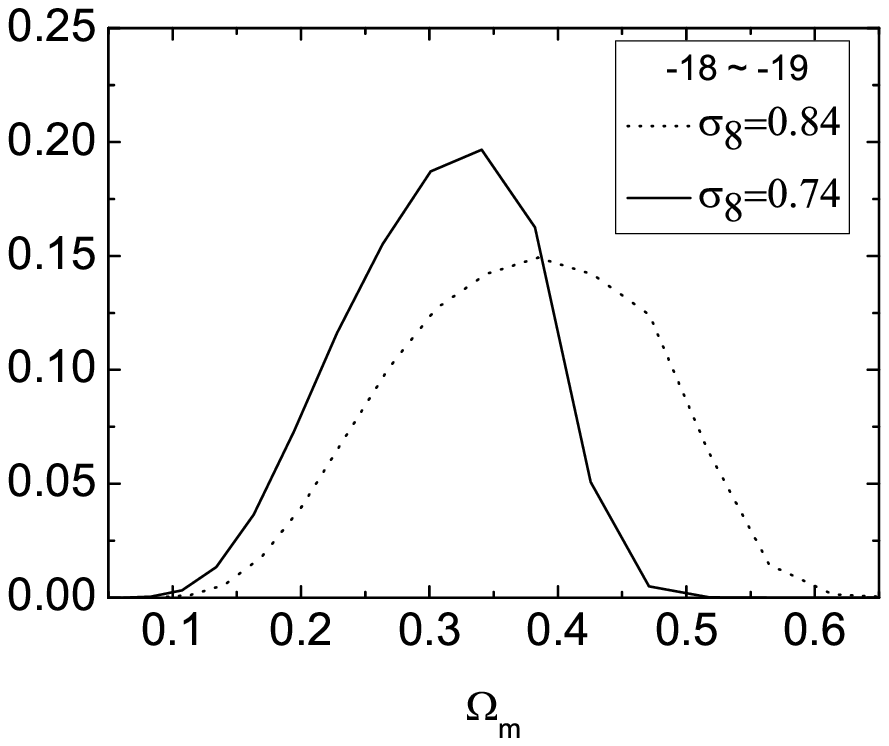}
\includegraphics{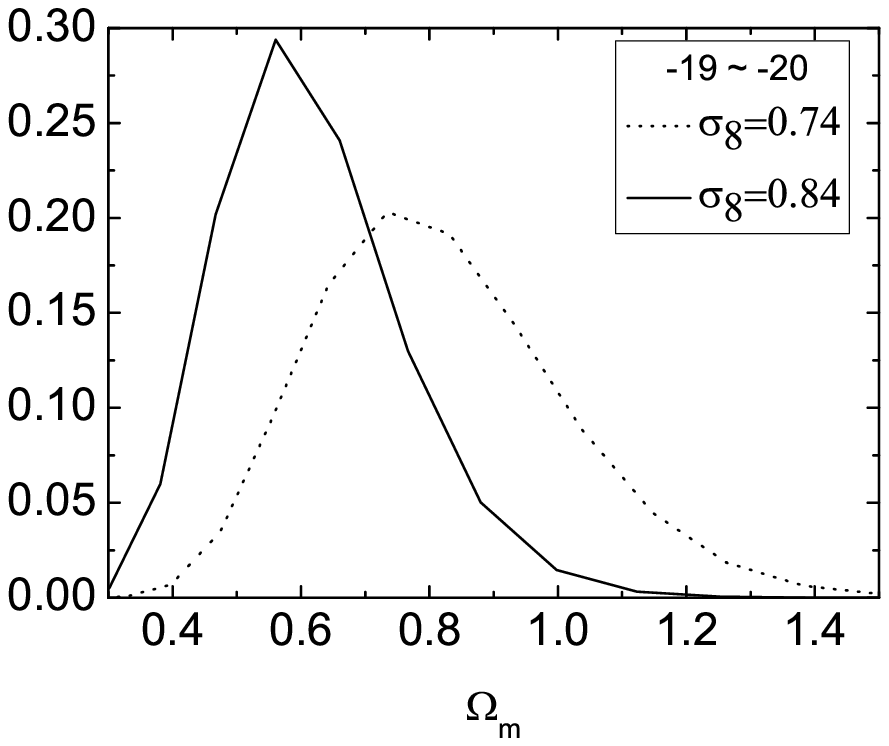}
\includegraphics{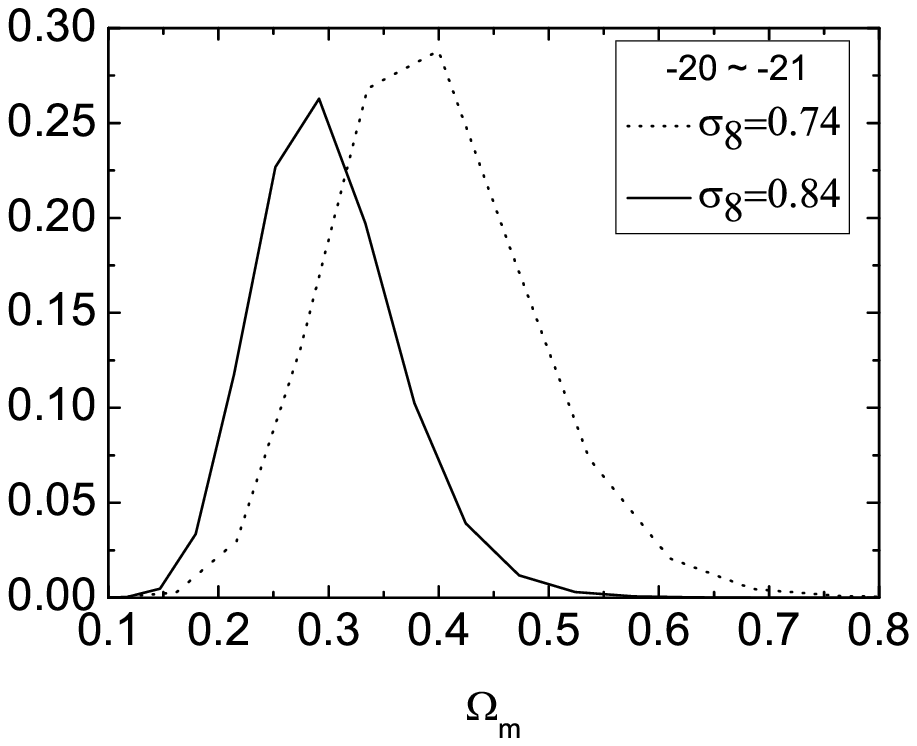}}
\caption{Marginalized distribution of $\Omega_m$ in three
parameter fitting with Eq.~(12). For all sub-samples, fitting
with lower $\sigma_8$ gives larger value of $\Omega_m$.
\label{fig-9}}
\end{figure*}

\begin{table*}
\centering \caption{Best parameters from fitting DWT power
spectra to $\Lambda\rm{CDM}$ model with $\sigma_8=0.84$ and
Eq.~(12).}
\begin{tabular}{lcccc}
\hline $M_{b_J}-5\log_{10} h$ & $\beta$ /$\Omega_m$ & $b$ & $\sigma_{pv}$ ($\rm{km s^{-1}}$) & $\chi^2$\\
\hline -19 --- -18  & $ 0.77^{+ 0.08}_{- 0.17} $
 /$0.34^{+0.15}_{-0.20}$ & $ 0.69^{+ 0.12}_{- 0.09}$
  & $315^{+ 68}_{- 44}$ &  10.91 \\
-20 --- -19  & $ 0.73^{+ 0.06}_{- 0.12} $ /$0.56^{+0.14}_{-0.22}$ &
$ 0.97^{+ 0.08}_{- 0.07}$
  & $578^{+ 50}_{- 40}$ &  8.46 \\
-21 --- -20  & $ 0.41^{+ 0.07}_{- 0.03} $ /$0.28^{+0.10}_{-0.05}$ &
$ 1.15^{+ 0.05}_{- 0.03}$
  & $548^{+ 68}_{- 26}$ &  4.96 \\
\hline
\end{tabular}
\label{tb:3p_high_sig8}
\end{table*}

\begin{table*}
\centering \caption{Best parameters from fitting DWT power
spectra to $\Lambda\rm{CDM}$ model with $\sigma_8=0.74$ and Eq.~(12).}
\begin{tabular}{lcccc}
\hline
$M_{b_J}-5\log_{10} h$ & $\beta$ /$\Omega_m$ & $b$ & $\sigma_{pv}$ ($\rm{km s^{-1}}$) & $\chi^2$\\
\hline
-19 --- -18  & $ 0.76^{+ 0.08}_{- 0.17} $ /$0.38^{+0.15}_{-0.20}$ & $ 0.74^{+ 0.10}_{- 0.07}$ & $336^{+ 57}_{- 41}$&  10.61 \\
-20 --- -19  & $ 0.73^{+ 0.06}_{- 0.12} $ /$0.73^{+0.19}_{-0.27}$ & $ 1.13^{+ 0.09}_{- 0.07}$ & $682^{+ 43}_{- 41}$&   8.26 \\
-21 --- -20  & $ 0.40^{+ 0.06}_{- 0.03} $ /$0.39^{+0.11}_{-0.07}$ & $ 1.42^{+ 0.03}_{- 0.06}$ & $1014^{+ 51}_{- 113}$&  4.89 \\
\hline
\end{tabular}
\label{tb:3p_low_sig8}
\end{table*}

\begin{table*}
\centering \caption{Best parameters $\sigma_8$, $b$, $\sigma_{pv}$ from fitting DWT
power spectra to  $\Lambda\rm{CDM}$ model with
$\Omega_m=0.24$, $\Omega_{\Lambda}=0.76$, and Eq.~(12).}
\begin{tabular}{lcccc}
\hline
$M_{b_J}-5\log_{10} h$ & $\sigma_8$ & $b$ & $\sigma_{pv}$$\rm{km s^{-1}}$ & $\chi^2$\\
\hline -19 --- -18 & $ 0.80^{+0.14}_{-0.08}$ & $0.73^{+ 0.21}_{-
0.12}$ &
   $325^{+66}_{-50}$&  17.00 \\
-20 --- -19  & $ 0.93^{+0.10}_{-0.07}$  & $ 0.91^{+ 0.15}_{-
0.14}$ &
   $470^{+57}_{-51}$&  15.62 \\
-21 --- -20  & $ 0.98^{+0.04}_{-0.04}$  & $ 0.97^{+ 0.05}_{-
0.04}$ &
   $412^{+50}_{-43}$&  7.16 \\
\hline
\end{tabular}
\label{tb:sig8_fit_Om_0.24}
\end{table*}

Empirically, the Fourier power spectrum $P^S(k)$ in redshift space is related to
that in real space by 
\citep{1994MNRAS.267.1020P}
%eq12
\begin{equation}
P^S(k)=b^2P(k)G(y,\beta)\ ,
\end{equation}
in which $y^2=k^2\sigma_{pv}^2$, and the function $G$ is
%eq13
\begin{equation}
\begin{aligned}
G(y,\beta)=&\frac{\sqrt{\pi}}{8}\frac{\rm{erf}(y)}{y^5}
[3\beta^2+4\beta y^2+4y^4]\\
&-\frac{\exp(-y^2)}{4y^4}[\beta^2(3+2y^2)+4\beta y^2]\ .
\end{aligned}
\end{equation}
Substituting Eq.(6) into Eq.(12) yields
%eq14
\begin{equation}
P^S_{j,j,j}(b,\beta,\sigma_{pv})=b^2P_{j,j,j}G[y(j),\beta]\ ,
\end{equation}
where  $y(j)=k^2(j)\sigma_{pv}^2$ and $k(j)$ is given by Eq.~(B12).

Following the same procedure in Section 5.2, we perform a three parameter ($b, \beta,
\sigma_{pv}$) fitting with Eq.(14). Fitting
results of model A and model B are written in Tables
\ref{tb:3p_high_sig8} and \ref{tb:3p_low_sig8} respectively.
PDFs of the three parameters $b, \beta,\sigma_{pv}$ for the sub-sample 
of $-21 \sim -20$ are plotted in Figure \ref{fig-8}, and the PDFs of $\Omega_m$ for 
all the three sub-samples are in \ref{fig-9}. Figure \ref{fig-8} shows that the models
A and B have very different PDF of $\sigma_{pv}$: the PDF of model B is highly skewed
and very broad while the PDF of model A is close to Gaussian.

The result of model A shown in Table \ref{tb:3p_high_sig8} is roughly the same as
that in Table \ref{tb:3p_DWT_sig0.84}. The sub-sample of $-19\sim -20$ which gives 
a large $\Omega_m$ but still agree with others within error bars. 
The results of model B shown in Table \ref{tb:3p_low_sig8} are similar to 
those in Table~\ref{tb:3p_DWT_sig0.74}. $\Omega_m$ from all sub-samples are larger 
than $0.3$, especially the one from the sub-sample of $-19\sim -20$ which is unusually
as large as $0.73^{+0.19}_{-0.27}$. Meanwhile, the pairwise velocity variance estimated 
from the sub-sample of $-21 \sim -20$ has an extraordinary value of  
$1014.28^{+ 51.42}_{- 113.57}\rm{km s^{-1}}$. These impose questions on the 
prior of $\sigma_8=0.74$.

Finally, we take $\sigma_8$, $b$ and $\sigma_{pv}$ as fitting
parameters, and fix $\Omega_m=0.24$. The best fitting parameters are 
in Table~\ref{tb:sig8_fit_Om_0.24}. Again we have the similar
results as those listed in Table \ref{tb:DWT_sig8_fit_Om_0.27}. All
the values of $\sigma_8$ are always $\geq 0.8$, which suggests that
the low $\sigma_8$ is unlikely to match with the 2dFGRS.

\section{Conclusion and Discussion}
DWT power spectra of 2dFGRS samples are measured on scales equivalent to
$0.04 <k< 2.3h$Mpc$^{-1}$. We show that these power spectra are
efficient to test $\Lambda$CDM models with high and low
amplitudes of mass density fluctuations. The model with $\sigma_8=0.84$ finds
good support from the 2dFGRS sample, all the
best fitting parameters, $b$, $\beta$, $\sigma_{pv}$ are consistent with
other works on 2dFGRS. Especially, three volume-limited samples gives
different $b$ and $\beta$, but approximately the same values of
$\Omega_m=(b\beta)^{1/0.6}= 0.28-0.36$. On the other hand, the model
with $\sigma_8=0.74$ cannot give such consistent fitting result, the
best fitted $\Omega_m$ from the three volume-limited
samples are significantly different, and in deviation from $\sim0.3$ at least at $1\sigma$
level. Moreover, the fitting results of $\sigma_{pv}$ are generically large, even reach
$10^3$ km s$^{-1}$. Our studies suggest that the power spectrum
of 2dFGRS disfavors models with low amplitude of
mass fluctuations, $\sigma_8=0.74$, if other cosmological parameters
are given by the WMAP3.

It is found that $\sigma_{pv}$ increases with luminosity, which is basically
consistent with the observation of \citet{2004ApJ...617..782J} though our estimations
are lower than theirs.

Another parameter that will affect the shape of the DWT power
spectrum is the slope of the primordial fluctuation spectrum $n_s$.
We have been using the scale-invariant spectrum, or the Zeldovich
spectrum with $n_s=1$. However, we notice that WMAP3 gives
$n_s=0.95^{+0.015}_{-0.019}$, which is 5\% smaller. In
order to check the influence of a lower value of $n_s$ on our fitting
results, we repeat our fitting for the sub-sample $-20 \sim -21$
with $n_s=0.95$. We find
$b=1.07^{+0.01}_{-0.01}$, $\beta=0.44^{+0.02}_{-0.02}$
($\Omega_m=0.29^{+0.02}_{-0.02}$), and
$\sigma_{pv}=550^{+20}_{-20}$km s$^{-1}$. Clearly, the change of
$n_s$ from 1 to 0.95 results in little modification to
$b$ and $\sigma_{pv}$, only slightly alters (less than 5\%) the
value of $\beta$, or $\Omega_m$.

In this paper, we used only the diagonal modes in term of vector
${\bf j}$, $P_{\bf j}=\langle \tilde{\epsilon}_{\bf j,l}
\tilde{\epsilon}_{\bf j,l}\rangle$, and ${\bf j}=(j,j,j)$. Even in the
second order statistics of the DWT, $\langle
\tilde{\epsilon}_{\bf j,l}\tilde{\epsilon}_{\bf j,l'}\rangle$, we
can have the power of off-diagonal modes $P_{\bf j}=\langle
\tilde{\epsilon}_{\bf j,l} \tilde{\epsilon}_{\bf j,l}\rangle$, and
${\bf j}=(j_1,j_2,j_3)$. In addition, we have the correlation between modes of
$({\bf j,l})$ and (${\bf j,l'})$ (${\bf l \neq l'}$). It has been
shown that different parts of the second order statistics of DWT
contains different information of the random field
\citep{2001PABei..19S..37Y,YangEtal2001a}. Possible constrains
on parameters given by various parts of the second order DWT
statistics deserve further study.

\section*{ACKNOWLEDGMENT}

We thank Shaun Cole, Carlton Baugh, Enzo Branchini, Steve Hatton and
their Durham astrophysics theory group and the VIRGO consortium for
the Hubble Volume mock catalogs. The Virgo $256^3$ simulation we
used in this paper was carried out by the Virgo Supercomputing
Consortium using computers based at Computing Centre of the
Max-Planck Society in Garching and at the Edinburgh Parallel
Computing Centre. The data are publicly available at
www.mpa-garching.mpg.de/NumCos. YCC and LLF are supported by NSFC under 
grant 10373012, JP is
supported by preliminary funding from PMO in the One-Hundred-Talent program
of CAS, and LZF is supported by the US NSF under grant AST-0507340.

%\bibliography{cc}

\begin{thebibliography}{}

\bibitem[\protect\citeauthoryear{{Bacon}, {Massey}, {Refregier} \&
  {Ellis}}{{Bacon} et~al.}{2003}]{2003MNRAS.344..673B}
{Bacon} D.~J.,  {Massey} R.~J.,  {Refregier} A.~R.,    {Ellis} R.~S.,  2003,
  \mnras, 344, 673

\bibitem[\protect\citeauthoryear{{Bahcall} \& {Bode}}{{Bahcall} \&
  {Bode}}{2003}]{2003ApJ...588L...1B}
{Bahcall} N.~A.,  {Bode} P.,  2003, \apjl, 588, L1

\bibitem[\protect\citeauthoryear{{Cole} et~al.,}{{Cole}
  et~al.}{2005}]{2005MNRAS.362..505C}
{Cole} S.,  et~al., 2005, \mnras, 362, 505

\bibitem[\protect\citeauthoryear{{Cole}, {Hatton}, {Weinberg} \&
  {Frenk}}{{Cole} et~al.}{1998}]{1998MNRAS.300..945C}
{Cole} S.,  {Hatton} S.,  {Weinberg} D.~H.,    {Frenk} C.~S.,  1998, \mnras,
  300, 945

\bibitem[\protect\citeauthoryear{{Colless} et~al.,}{{Colless}
  et~al.}{2001}]{2001MNRAS.328.1039C}
{Colless} M.,  et~al., 2001, \mnras, 328, 1039

\bibitem[\protect\citeauthoryear{{Colless} et~al.,}{{Colless}
  et~al.}{2003}]{2003yCat.7226....0C}
{Colless} M.,  et~al., 2003, VizieR Online Data Catalog, 7226, 0

\bibitem[\protect\citeauthoryear{{Daubechies}}{{Daubechies}}{1992}]{1992DauTLO%
W}
{Daubechies} I.,  1992, Ten Lectures on Wavelets. Philadelphia, SIAM.

\bibitem[\protect\citeauthoryear{{Fang} \& {Feng}}{{Fang} \&
  {Feng}}{2000}]{2000ApJ...539....5F}
{Fang} L.-Z.,  {Feng} L.-l.,  2000, \apj, 539, 5

\bibitem[\protect\citeauthoryear{{Fang} \& {Thews}}{{Fang} \&
  {Thews}}{1998}]{1998Fang}
{Fang} L.-Z.,  {Thews} R.,  1998, Wavelet in Physics. World Scientific,
  Singapore.

\bibitem[\protect\citeauthoryear{{Hawkins} et~al.,}{{Hawkins}
  et~al.}{2003}]{2003MNRAS.346...78H}
{Hawkins} E.,  et~al., 2003, \mnras, 346, 78

\bibitem[\protect\citeauthoryear{{Hoekstra}, {Yee} \& {Gladders}}{{Hoekstra}
  et~al.}{2002}]{2002ApJ...577..595H}
{Hoekstra} H.,  {Yee} H.~K.~C.,    {Gladders} M.~D.,  2002, \apj, 577, 595

\bibitem[\protect\citeauthoryear{{Jing} \& {B{\"o}rner}}{{Jing} \&
  {B{\"o}rner}}{2004}]{2004ApJ...617..782J}
{Jing} Y.~P.,  {B{\"o}rner} G.,  2004, \apj, 617, 782

\bibitem[\protect\citeauthoryear{{Mallat}}{{Mallat}}{1989a}]{1989Mallata}
{Mallat} S.~G.,  1989a, IEEE Trans, on PAMI, 11, 674

\bibitem[\protect\citeauthoryear{{Mallat}}{{Mallat}}{1989b}]{1989Mallatb}
{Mallat} S.~G.,  1989b, Trans. Am. Math. Soc. 315, 69

\bibitem[\protect\citeauthoryear{{Meyer}}{{Meyer}}{1992}]{1992Meyer}
{Meyer} Y.,  1992, Wavelets and Operators. Cambridge Press, New York.

\bibitem[\protect\citeauthoryear{{Norberg} et~al.,}{{Norberg}
  et~al.}{2001}]{2001MNRAS.328...64N}
{Norberg} P.,  et~al., 2001, \mnras, 328, 64

\bibitem[\protect\citeauthoryear{{Norberg} et~al.,}{{Norberg}
  et~al.}{2002}]{2002MNRAS.336..907N}
{Norberg} P.,  et~al., 2002, \mnras, 336, 907

\bibitem[\protect\citeauthoryear{{Pan} \& {Szapudi}}{{Pan} \&
  {Szapudi}}{2005}]{2005MNRAS.362.1363P}
{Pan} J.,  {Szapudi} I.,  2005, \mnras, 362, 1363

\bibitem[\protect\citeauthoryear{{Pando} \& {Fang}}{{Pando} \&
  {Fang}}{1995}]{1995AAS...187.8502P}
{Pando} J.,  {Fang} L.-Z.,  1995, Bulletin of the American Astronomical
  Society, 27, 1411

\bibitem[\protect\citeauthoryear{{Pando} \& {Fang}}{{Pando} \&
  {Fang}}{1996}]{1996ApJ...459....1P}
{Pando} J.,  {Fang} L.-Z.,  1996, \apj, 459, 1

\bibitem[\protect\citeauthoryear{{Peacock} \& {Dodds}}{{Peacock} \&
  {Dodds}}{1994}]{1994MNRAS.267.1020P}
{Peacock} J.~A.,  {Dodds} S.~J.,  1994, \mnras, 267, 1020

\bibitem[\protect\citeauthoryear{{Peacock} et~al.,}{{Peacock}
  et~al.}{2001}]{2001Natur.410..169P}
{Peacock} J.~A.,  et~al., 2001, \nat, 410, 169

\bibitem[\protect\citeauthoryear{{Percival} et~al.,}{{Percival}
  et~al.}{2001}]{2001MNRAS}
{Percival} W.~J.,  et~al., 2001, \mnras, 327, 1297

\bibitem[\protect\citeauthoryear{{Percival} et~al.,}{{Percival}
  et~al.}{2004}]{2004MNRAS.353.1201P}
{Percival} W.~J.,  et~al., 2004, \mnras, 353, 1201

\bibitem[\protect\citeauthoryear{{Refregier}, {Rhodes} \& {Groth}}{{Refregier}
  et~al.}{2002}]{2002ApJ...572L.131R}
{Refregier} A.,  {Rhodes} J.,    {Groth} E.~J.,  2002, \apjl, 572, L131

\bibitem[\protect\citeauthoryear{{Reiprich} \& {B{\"o}hringer}}{{Reiprich} \&
  {B{\"o}hringer}}{2002}]{2002ApJ...567..716R}
{Reiprich} T.~H.,  {B{\"o}hringer} H.,  2002, \apj, 567, 716

\bibitem[\protect\citeauthoryear{{Seljak} et~al.,}{{Seljak}
  et~al.}{2005}]{2005PhRvD..71j3515S}
{Seljak} U.,  et~al., 2005, \prd, 71, 103515

\bibitem[\protect\citeauthoryear{{Smith}, {Peacock}, {Jenkins}, {White},
  {Frenk}, {Pearce}, {Thomas}, {Efstathiou} \& {Couchman}}{{Smith}
  et~al.}{2003}]{2003MNRAS.341.1311S}
{Smith} R.~E.,  {Peacock} J.~A.,  {Jenkins} A.,  {White} S.~D.~M.,  {Frenk}
  C.~S.,  {Pearce} F.~R.,  {Thomas} P.~A.,  {Efstathiou} G.,    {Couchman}
  H.~M.~P.,  2003, \mnras, 341, 1311

\bibitem[\protect\citeauthoryear{{Spergel}, {Bean}, {Nolta} \&
  {Bennett}}{{Spergel} et~al.}{2006}]{Spergel2006}
{Spergel} D.~N.,  {Bean} R.,  {Nolta} M.~R.,    {Bennett} C.~L.,  2006,
  astro-ph/0603449

\bibitem[\protect\citeauthoryear{{Tegmark} et~al.,}{{Tegmark}
  et~al.}{2004}]{2004PhRvD..69j3501T}
{Tegmark} M.,  et~al., 2004, \prd, 69, 103501

\bibitem[\protect\citeauthoryear{{Tegmark}, {Hamilton} \& {Xu}}{{Tegmark}
  et~al.}{2002}]{2002MNRAS.335..887T}
{Tegmark} M.,  {Hamilton} A.~J.~S.,    {Xu} Y.,  2002, \mnras, 335, 887

\bibitem[\protect\citeauthoryear{{Van Waerbeke}, {Mellier}, {Pell{\'o}}, {Pen},
  {McCracken} \& {Jain}}{{Van Waerbeke} et~al.}{2002}]{2002A&A...393..369V}
{Van Waerbeke} L.,  {Mellier} Y.,  {Pell{\'o}} R.,  {Pen} U.-L.,  {McCracken}
  H.~J.,    {Jain} B.,  2002, \aap, 393, 369

\bibitem[\protect\citeauthoryear{{Viel} \& {Haehnelt}}{{Viel} \&
  {Haehnelt}}{2006}]{2006MNRAS.365..231V}
{Viel} M.,  {Haehnelt} M.~G.,  2006, \mnras, 365, 231

\bibitem[\protect\citeauthoryear{{Yang}, {Feng} \& {Chu}}{{Yang}
  et~al.}{2001}]{2001PABei..19S..37Y}
{Yang} X.,  {Feng} L.,    {Chu} Y.,  2001, Progress in Astronomy, 19, 37

\bibitem[\protect\citeauthoryear{{Yang}, {Feng}, {Chu} \& {Fang}}{{Yang}
  et~al.}{2001a}]{YangEtal2001a}
{Yang} X.,  {Feng} L.-L.,  {Chu} Y.,    {Fang} L.-Z.,  2001a, \apj, 553, 1

\bibitem[\protect\citeauthoryear{{Yang}, {Feng}, {Chu} \& {Fang}}{{Yang}
  et~al.}{2001b}]{2001ApJ...560..549Y}
{Yang} X.,  {Feng} L.-L.,  {Chu} Y.,    {Fang} L.-Z.,  2001b, \apj, 560, 549

\bibitem[\protect\citeauthoryear{{Yang}, {Feng}, {Chu} \& {Fang}}{{Yang}
  et~al.}{2002}]{2002ApJ...566..630Y}
{Yang} X.,  {Feng} L.-L.,  {Chu} Y.,    {Fang} L.-Z.,  2002, \apj, 566, 630

\bibitem[\protect\citeauthoryear{{Zhan} \& {Fang}}{{Zhan} \&
  {Fang}}{2003}]{2003ApJ...585...12Z}
{Zhan} H.,  {Fang} L.-Z.,  2003, \apj, 585, 12

\end{thebibliography}
%\bibliographystyle{mn2e}

\appendix

\section{DWT decomposition of random field}

For the details of the mathematical properties of the DWT, please
refer to
\citet{1989Mallata,1989Mallatb,1992Meyer,1992DauTLOW}, and for
physical applications, refer to \citet{1998Fang}. For this application,
the most important properties are 1.) orthogonality, 2.)
completeness, and 3.) locality in both scale ($r$) and physical
position ($x$). Wavelets with compactly supported basis are an
excellent means to analyze random fields. Among the compactly
supported orthogonal wavelets, the Daubechies family of wavelets
are easy to implement.

To simplify the notation, we consider a 1-D field $\rho(x)$ on
spatial range $L$. It is straightforward to generalize to 3-D
fields. In DWT analysis, the space $L$ is chopped into $2^j$
segments labeled by $l=0,1,...2^j-1$. Each of the segments has
size $L/2^j$. The index $j$ is a positive integer which represents
scale $L/2^j$. The index $l$ gives position and corresponds to
spatial range $lL/2^j < x < (l+1)L/2^j$.

DWT analysis uses two functions, the scaling functions
$\phi_{j,l}(x)=(2^j/L)^{1/2}\phi(2^j/L-l)$, and wavelets
$\psi_{j,l}(x)=(2^j/L)^{1/2}\psi(2^j/L-l)$. The scaling functions
and wavelets  are given, respectively, by a translation and
dilation of the basic scaling function $\phi(\eta)$ and basic
wavelet $\psi(\eta)$ as
%A1
\begin{equation}
\phi_{j,l}(x) = \left (\frac{2^{j}}{L}\right)^{1/2} \phi(2^jx/L -
l)
\end{equation}
and
%A2
\begin{equation}
\psi_{j,l}(x) = \left (\frac{2^{j}}{L}\right)^{1/2}\psi(2^jx/L-l).
\end{equation}
The scaling functions play the role of window function.  They are
used to calculate the mean field in the segment $l$. The wavelets
$\psi_{j,l}(x)$ capture the difference between the mean fields at
space ranges $lL/2^j < x < (l+1/2)L/2^j$ and $(l+1/2)L/2^j < x <
(l+1)L/2^j$.

The scaling functions and wavelets $\psi_{j,l}(x)$ satisfy the
orthogonal relations
%eqA3
\begin{equation}
\int \phi_{j,l}(x)\phi_{j,l'}(x)dx= \delta_{l,l'},
\end{equation}
%eqA4
\begin{equation}
\int \psi_{j,l}(x)\psi_{j',l'}(x)dx=\delta_{j,j'} \delta_{l,l'},
\end{equation}
%eqA5
\begin{equation}
\int\phi_{j,l}(x)\psi_{j',l'}(x)dx =0, \ \ \ \mbox{if $j'\geq j$}.
\end{equation}

With these properties, a 1-D random field $\rho(x)$ can be
decomposed into
%eqA6
\begin{equation}
\rho(x) = \rho^{j}(x) + \sum_{j'=j}^{\infty} \sum_{l=0}^{2^{j'}-1}
  \tilde{\epsilon}_{j',l} \psi_{j',l}(x),
\end{equation}
where
%eqA7
\begin{equation}
\rho^j(x)=\sum_{l=0}^{2^j-1}\epsilon_{j,l}\phi_{j,l}(x).
\end{equation}
The scaling function coefficient (SFC) $\epsilon_{j,l}$ and the
wavelet function coefficient (WFC), $\tilde{\epsilon}_{j,l}$ are
given by
%eqA8
\begin{equation}
\epsilon_{j,l} =\int \rho(x)\phi_{j,l}(x)dx,
\end{equation}
and
%eqA9
\begin{equation}
\tilde{\epsilon}_{j,l}=\int \rho(x)\psi_{j,l}(x)dx,
\end{equation}
respectively. The SFC $\epsilon_{j,l}$ measure the mean of
$\rho(x)$ in the segment $l$, while the WFC
$\tilde{\epsilon}_{j,l}$ measures the fluctuations (or difference)
of field $\rho(x)$ at $l$ on scale $j$.

The first term on the r.h.s. of Eq.(A6), $\rho^{j}(x)$, is the
field $\rho(x)$ smoothed on the scale $j$, while the second term
contains all information on scales $\geq j$. Because of the
orthogonality, the decomposition between the scales of $ <j$
(first term) and $\geq j$ (second term) in Eq.(A6) is unambiguous.

\section{1-D DWT Power Spectrum}

The contrast (or perturbation) of the field $\rho(x)$ is defined
by
%eqB1
\begin{equation}
\delta(x) = \frac{\rho(x) - \bar\rho}{\bar\rho}
\end{equation}
where $\bar\rho$ is the mean density of the field. The Fourier
expansion of $\epsilon$ is
%eqB2
\begin{equation}
\delta (x)=\sum_{n = - \infty}^{\infty} \epsilon_n e^{i2\pi nx/L}
\end{equation}
with the coefficients given by
%eqB3
\begin{equation}
\epsilon_n= \frac{1}{L}\int_0^{L} \delta(x)e^{-i2\pi nx/L}dx.
\end{equation}

Parseval's theorem relates the power for a distribution to the
coefficients of the Fourier expansion. For the contrast this
yields
%eqB4
\begin{equation}
\frac{1}{L} \int_0^L |\epsilon(x)|^2 dx    \sum_{n= -\infty}^{\infty} |\epsilon_n|^2,
\end{equation}
which shows that the perturbations can be decomposed into domains,
$n$, by the orthonormal Fourier basis functions. The power
spectrum of perturbations  on scale $L/n$ is then defined as
%eqB5
\begin{equation}
P(n)= |\epsilon_n|^2.
\end{equation}
This is the power spectrum with respect to the Fourier
decomposition.

Similarly Parseval's theorem for the DWT is
%eqB6
\begin{equation}
\frac{1}{L}\int_0^L |\epsilon(x)|^2 dx = \sum_{j= 0}^{\infty}
\frac{1}{L}\sum_{l=0}^{2^j-1} |\tilde{\epsilon}_{j,l}|^2.
\end{equation}
Thus, the second order statistical behavior of $\epsilon(x)$ can
be described by the $|\tilde{\epsilon}_{j,l}|^2$ and one can call
$|\tilde{\epsilon}_{j,l}|^2$  the DWT power spectrum.

Comparing Eqs.(B4) and (B6), it is clear that
$\frac{1}{L}\sum_{l=0}^{2^j-1}|\tilde{\epsilon}_{j,l}|^2$ is a
measure of the power of the perturbation on scales from $L/2^j$
to $L/2^{j+1}$. Therefore, the power spectrum with
respect to the wavelet basis can be defined as
%eqB7
\begin{equation}
P_j =\frac{1}{2^j}\sum_{l=0}^{2^j-1} |\tilde{\epsilon}_{j,l}|^2.
\end{equation}
Since the DWT bases $\psi_{j,l}(x)$ measure the differences
between the {\it local} mean densities at adjoining scales, the
mean density on length scales larger than the sample size is not
needed in calculating $\tilde{\epsilon}_{j,l}$. The spectrum
Eq.(B7) will not be affected by the infrared (long-wavelength)
uncertainty of the mean density.

With Eqs.(B2), (B6), (A2) and (A4) we can find the relation
between the power spectra of DWT $P_j$ and Fourier $P(n)$. It is
%eqB8
\begin{equation}
P_j = \frac{1}{2^j} \sum_{n = - \infty}^{\infty}
 |\hat{\psi}(n/2^j)|^2 P(n),
\end{equation}
where $\hat{\psi}(n)$ is the Fourier transform of the basic
wavelet given by
%eqB9
\begin{equation}
\hat{\psi}(n)= \int_{0}^{L} \psi(\eta) e^{-i2\pi n\eta}d\eta.
\end{equation}
Since wavelet is admissible, i.e. $\int \psi(\eta)d\eta=0$,
we have $\hat{\psi}(0)=0$. $|\hat{\psi}(n)|^2$ is localized in
$n$-space. $|\hat{\psi}(n)|^2$ has symmetrically
distributed peaks with respect to $n=0$. The first highest peaks
are non-zero in two narrow ranges centered at $n=\pm n_p$ with
width $\Delta n_p$. Besides the first peak, there are ``side
lobes" in $|\hat{\psi}(n)|^2$. However, the ``side lobes" is small,
for instance, for the Daubechies 4 wavelet, the area under the
``side lobes" is not more than 2\% of the first peak.
Therefore, $P_j$ is a good estimation of the band-averaged
Fourier power spectrum centered at wavenumber
%eqB10
\begin{equation}
 n_j = n_p 2^j.
\label{eq11}
\end{equation}
The band width is
%eqB11
\begin{equation}
\Delta n =2^j \Delta n_p. \label{eq12}
\end{equation}
In other words, the relation between k and j is
\begin{equation}
\log k=(\log 2)j-\log(L/2\pi)+\log n_p.
\end{equation}
For the D4 wavelet, $\log n_p=0.270$.

\section{3-D DWT Power Spectrum}

For 3-D random field, the DWT decomposition is based on the
orthogonal and complete set of 3-D wavelet basis $\{\psi_{\bf
j,l}({\bf x})\}$, which can be constructed by a direct product
of 1-D wavelet basis as
%eqC1
\begin{equation}
\psi_{\bf j,l}({\bf x})=\psi_{j_1,l_1}(x_1)\psi_{j_2,l_2}(x_2)
 \psi_{j_3,l_3}(x_3),
\end{equation}
where $j_i=0,1,2..$ ($i=1,2,3$) and $l_i=0...2^{j_i-1}$. Obviously,
the basis $\psi_{\bf j,l}({\bf x})$ is non-zero mainly in a volume
$L_1/2^{j_1}\times L_2/2^{j_2}\times L_3/2^{j_3}$, and  around the
position $(x_1=l_1L_1/2^{j_1}, x_2=l_2L_2/2^{j_2},x_3=l_3L_3/2^{j_3})$
\citep{1998Fang}

Similar to Eq.(B7), the power spectrum on scale ${\bf j}=(j_1,j_2,j_3)$
is
%eqC2
\begin{equation}
P_{\bf j}^2= \frac{1}{2^{j_1+j_2+j_3}}\sum_{l_1=0}^{2^{j_1}-1}
\sum_{l_2=0}^{2^{j_2}-1}\sum_{l_3=0}^{2^{j_3}-1}
  [\tilde\epsilon_{\bf j,l}]^2,
\end{equation}
For 3-D samples, Eq.(B8) is generalized as
%eqC3
\begin{equation}
\begin{aligned}
P_{\bf j} = &\frac{1}{2^{j_1+j_2+j_3}}
  \sum_{n_1 = - \infty}^{\infty}
  \sum_{n_2 = - \infty}^{\infty}
  \sum_{n_3 = - \infty}^{\infty}\\
  &|\hat{\psi}(n_1/2^{j_1})\hat{\psi}(n_2/2^{j_2})
  \hat{\psi}(n_3/2^{j_3})|^2 P(n_1,n_2,n_3).
\end{aligned}
\end{equation}
Because the cosmic density field is isotropic, the Fourier power
spectrum $P(n_1,n_2,n_3)$ is dependent only on
%eqC4
\begin{equation}
n=\sqrt{n_1^2+n_2^2+n_3^2}.
\end{equation}
Obviously, the DWT power spectrum is invariant with respect to the
cyclic permutation of index as
%eqC5
\begin{equation}
P_{j_1,j_2,j_3}=P_{j_3,j_1,j_2}=P_{j_2,j_3,j_1}
\end{equation}

Considering Eq.(C3) and Eq.(C4), we can formally define
a band center wavenumber $n_j$ corresponding to the 3-D mode ${\bf
j}$ as
%eqC6
\begin{equation}
n_{\bf j} = n_p \sqrt{(2^{j_1})^2+(2^{j_2})^2+(2^{j_3})^2}.
\end{equation}
For an isotropic random field, the Fourier modes with the same $n$
are statistically equivalent. However, the DWT modes
with the same $n_j$ [Eq.(C6)] are not statistically
equivalent, because the DWT modes are not rotationally invariant. A
Fourier mode $e^{-i(2\pi/L)(n_1x_1+n_2x_2 +n_3x_3)}$ can be obtained
by a rotation of mode $e^{-i(2\pi/L)(n'_1x_1+ n'_2x_2 +n'_3x_3)}$ as
long as $n'^2_1+n'^2_2+n'^2_3 =n_1^2 + n_2^2 + n_3^2$. However, the
DWT modes don't have the same property. Generally, one cannot
transform a mode $(j_1, j_2, j_3)$ to $(j'_1, j'_2, j'_3)$ by a
rotation, even when $n_j \simeq n_{j'}$. Because of different
configurations between them, the condition $n_j n_{j'}$ generally does not imply
%eqC7
\begin{equation}
P_{\bf j}=P_{\bf j'}.
\end{equation}
This invariance holds only when $(j_1, j_2, j_3)$ is a cyclic
permutation of $(j'_1, j'_2, j'_3)$.

With this property, one can define two types of the DWT power
spectra: 1. the diagonal power spectrum given by $P_{\bf j}$ on
diagonal modes $j_1=j_2=j_3=j$, and 2. off-diagonal power spectrum
given by other modes.

From Eq.(C3), the diagonal power spectrum $P_j \equiv P_{j,j,j}$
is related to the Fourier power spectrum by
%eqC8
\begin{equation}
P_j   \sum_{n_1 = - \infty}^{\infty}
  \sum_{n_2 = - \infty}^{\infty}
  \sum_{n_3 = - \infty}^{\infty}W_j(n_1,n_2,n_3)
   P(n_1,n_2,n_3),
\end{equation}
where the window function $W_j$ is
\begin{equation}
W_j(n_1,n_2,n_3)\frac{1}{2^{3j}}|\hat{\psi}(n_1/2^{j})\hat{\psi}(n_2/2^{j})
  \hat{\psi}(n_3/2^{j})|^2.
\end{equation}
with the normalization
\begin{equation}
\int_{-\infty}^{\infty} W_j(n_1,n_2,n_3) dn_1dn_2dn_3=1.
\end{equation}
The window function $W_j$ is localized around $n_1=n_2=n_3= n_p2^j$.
Therefore, the diagonal power spectrum $P_j$ is a band-average
of the isotropic Fourier power spectrum $P(n)$ with the central
frequency $n=\sqrt{3}n_g2^j$.

There are two types of modes: diagonal mode with $j_1=j_2=j_3$;
off-diagonal mode for which the three numbers $(j_1,j_2,j_3)$ are not
the same. The DWT estimator can provide two types of power spectra:
1. diagonal power spectrum given by the powers on diagonal modes,
2. off-diagonal power spectrum given by the powers on off-diagonal
modes. Because the two types of modes have different spatial invariance,
the diagonal and off-diagonal DWT power spectra are very flexible
to deal with configuration-related problems in the power spectrum
detection. For off-diagonal modes, one can also calculate the linear
non-diagonal DWT power spectrum $P_{j_1,j_2,j_3}$ via Eq.(C2).
However, in this case, $P_{j_1,j_2,j_3}$ cannot simply be identified
as a band average of the isotropic Fourier power spectrum $P(n)$
centered at $n=n_j$. Nevertheless, $n_j$ is useful to calibrate the
physical scale of a given ${\bf j}$.

\end{document}